\documentclass[journal]{IEEEtran}
\usepackage{amsmath,amsfonts}
\usepackage{algorithmic}
\usepackage{algorithm}
\usepackage{array}
\usepackage[caption=false,font=normalsize,labelfont=sf,textfont=sf]{subfig}
\usepackage{textcomp}
\usepackage{stfloats}
\usepackage{url}
\usepackage{verbatim}
\usepackage{graphicx}
\usepackage{cite}

\usepackage{amssymb}
\usepackage{booktabs}       
\usepackage{multirow}       
\usepackage{siunitx}        
\usepackage{threeparttable} 
\usepackage[hidelinks]{hyperref}       
\usepackage{xspace}         
\usepackage{color}          
\usepackage{pifont}          
\usepackage{microtype}       
\usepackage{enumitem}        
\usepackage[most]{tcolorbox} 
\usepackage{balance}         

\graphicspath{ {./figures/}{./} } 

\newcommand{\RQone}{Which repository-level activity and responsiveness signals most reliably distinguish sustained maintenance from long-term maintenance decline across package ecosystems?\xspace}
\newcommand{\RQtwo}{How does Time Lag vary across maintenance states, and to what extent do packages with low Version Lag still exhibit signs of maintenance decline?\xspace}
\newcommand{\RQthree}{To what extent does incorporating maintenance decline into TL models change the identification and ranking of high-risk dependencies compared to Version Lag measures?\xspace}
\newcommand{\DataShare}{\url{https://figshare.com/s/a1cd42af90b13e09d61c}\xspace}

\newcommand{\rqoneNPackages}{11,047}
\newcommand{\rqoneNArchived}{628}
\newcommand{\rqonePctArchived}{5.7\%}
\newcommand{\rqoneDASMean}{0.313}
\newcommand{\rqoneDASMedian}{0.114}
\newcommand{\rqoneMRSMedian}{0.498}
\newcommand{\rqoneRMVSMean}{0.538}
\newcommand{\rqoneRMVSMedian}{0.547}
\newcommand{\rqoneFinalMedian}{0.235}
\newcommand{\rqoneVeryActiveDASMedian}{0.704}
\newcommand{\rqoneVeryActiveRMVSMedian}{0.605}
\newcommand{\rqoneVeryActiveFinalMedian}{0.616}
\newcommand{\rqoneSedentaryDASMedian}{0.000}
\newcommand{\rqoneSedentaryRMVSMedian}{0.517}
\newcommand{\rqoneSedentaryFinalMedian}{0.086}

\newcommand{\rqoneNActive}{3,307}
\newcommand{\rqoneNDeclining}{7,740}
\newcommand{\rqoneNMRSDefined}{6,155}
\newcommand{\rqoneDASAUCActive}{0.803}

\newcommand{\rqoneMRSAUCActive}{0.550}

\newcommand{\rqoneRMVSAUCActive}{0.644}
\newcommand{\rqoneRMVSAUCArchived}{0.999}
\newcommand{\rqoneFinalAUCActive}{0.783}
\newcommand{\rqoneFinalAUCArchived}{0.944}

\newcommand{\rqtwoNPackages}{11,047}
\newcommand{\rqtwoNCapped}{3,056}
\newcommand{\rqtwoCappedPct}{27.7}
\newcommand{\rqtwoTimeLagCapDays}{3,652}
\newcommand{\rqtwoLowVNDThreshold}{1.0}

\newcommand{\rqtwoVATimeLagMedian}{25}
\newcommand{\rqtwoVALongPct}{8.4}
\newcommand{\rqtwoMATimeLagMedian}{45}
\newcommand{\rqtwoMALongPct}{12.2}
\newcommand{\rqtwoLATimeLagMedian}{115}
\newcommand{\rqtwoLALongPct}{20.8}
\newcommand{\rqtwoSedTimeLagMedian}{879}
\newcommand{\rqtwoSedLongPct}{66.2}
\newcommand{\rqtwoTimeLagMedianRatio}{35.2$\times$}

\newcommand{\rqtwoLongTermRatio}{7.9$\times$}

\newcommand{\rqtwoLowVLN}{9,918}

\newcommand{\rqtwoLowVLLowMaltaPct}{21.6}
\newcommand{\rqtwoLowVLLowMaltaTL}{20}

\newcommand{\rqtwoLowVLMedMaltaPct}{16.2}

\newcommand{\rqtwoLowVLHighMaltaN}{6,167}
\newcommand{\rqtwoLowVLHighMaltaPct}{62.2}
\newcommand{\rqtwoLowVLHighMaltaTL}{1067}
\newcommand{\rqtwoLowVLHighMaltaArchPct}{9.8}
\newcommand{\rqtwoLowVLHiddenRatio}{53.4$\times$}
\newcommand{\rqtwoAllHighMaltaPct}{57.7}

\newcommand{\rqtwoLowVLAbandonedCombPct}{62.2}

\newcommand{\rqthreeNPackages}{11,047}

\newcommand{\rqthreeLagLowTotal}{9,918}

\newcommand{\rqthreeLagLowMaltaLowPct}{21.6}

\newcommand{\rqthreeLagLowMaltaHighN}{6,167}
\newcommand{\rqthreeLagLowMaltaHighPct}{62.2}

\newcommand{\rqthreeLagHighMaltaLowPct}{66.7}
\newcommand{\rqthreeChiSq}{936.61}
\newcommand{\rqthreeCramersV}{0.206}
\newcommand{\rqthreeFalseHealthyN}{6,167}

\newcommand{\rqthreeFalseHealthyMALTAMean}{0.109}
\newcommand{\rqthreeFalseHealthyTimeLagMean}{2019}
\newcommand{\rqthreeFalseHealthyArchivedPct}{9.8}
\newcommand{\rqthreeFalseHealthySedentaryPct}{81.8}
\newcommand{\rqthreeTrulyHealthyN}{2,143}
\newcommand{\rqthreeTrulyHealthyMALTAMean}{0.786}
\newcommand{\rqthreeTrulyHealthyTimeLagMean}{36}
\newcommand{\rqthreeTrulyHealthyArchivedPct}{0.0}
\newcommand{\rqthreeTimeLagRatio}{55.5}

\newcommand{\rqthreeLagHighMaltaHighPct}{17.7}

\newcommand{\rqthreeMaltaHighRateRatio}{3.5}

\newcommand{\rqoneAUCNoArchFinal}{0.686}

\newcommand{\rqoneArchHighRiskPct}{90.9}

\newcommand{\rqoneArchLowRiskPct}{2.9}
\newcommand{\rqoneNonArchHighRiskPct}{55.3}

\begin{document}
\title{MALTA: Maintenance-Aware Technical Lag, Estimation to Address Software Abandonment}
\author{Shane K. Panter and Nasir U. Eisty
    \thanks{Shane K. Panter is with Boise State University, Boise, ID, USA (email: shanepanter@boisestate.edu). Nasir U. Eisty is with The University of Tennessee, Knoxville, TN, USA (email: neisty@utk.edu).}
}
\markboth{IEEE Transactions on Software Engineering}%
{Panter and Eisty: On the relationship between Technical Lag and Software Abandonment}

\maketitle
\begin{abstract}
\textbf{Context:} Open-source ecosystems rely on sustained package maintenance. When maintenance slows or stops, Technical Lag (TL), the gap between installed and latest dependency versions accumulates, creating security and sustainability risks. However, some existing TL metrics, such as Version Lag, struggle to distinguish between actively maintained and abandoned packages, leading to a systematic underestimation of risk.
\textbf{Objective:} We investigate the relationship between Version Lag and software abandonment by (i) identifying which repository-level signals reliably distinguish sustained maintenance from long-term decline, (ii) quantifying how Version Lag magnitude and persistence differ across maintenance states, and (iii) evaluating how maintenance-aware metrics change the identification of high-risk dependencies.
\textbf{Method:} We introduce Maintenance-Aware Lag and Technical Abandonment (MALTA), a scoring framework comprising three metrics: Development Activity Score (DAS), Maintainer Responsiveness Score (MRS), and Repository Metadata Viability Score (RMVS). We evaluate MALTA on a dataset of \rqoneNPackages{} Debian packages linked to upstream GitHub repositories, encompassing 1.7 million commits and 4.2 million pull requests.
\textbf{Results:} MALTA achieves AUC = \rqoneFinalAUCActive{} for classifying active versus declining maintenance. Most significantly, \rqthreeLagLowMaltaHighPct{}\% of packages classified as ``Low Risk'' by Version Lag alone are reclassified as ``High Risk'' when MALTA signals are incorporated. These discordant packages average \rqthreeFalseHealthyTimeLagMean{} days since their last commit, with \rqthreeFalseHealthyArchivedPct{}\% having archived repositories.
\textbf{Conclusions:} Version Lag metrics systematically miss abandoned packages, a blind spot affecting the majority of dependencies in distribution ecosystems. MALTA identifies a substantial discordant population invisible to Version Lag by distinguishing \textit{resolvable lag} from \textit{terminal lag} caused by upstream abandonment.
\end{abstract}


\section{Introduction}\label{sec:intro}

Technical lag (TL) is defined as the delay in updating software dependencies to their latest available versions~\cite{gonzalez-barahonaTechnicalLagSoftware2017}. It is a hidden risk that can lead to various issues, including security vulnerabilities~\cite{cox_measuring_2015}, compatibility problems, and increased maintenance costs. Managing TL is crucial for ensuring the health and sustainability of software projects, particularly in open-source ecosystems where dependencies are often numerous and rapidly evolving.

Prior work~\cite{panterTechnicalLagLatent2026a} consolidated existing research and mapped existing metrics into six distinct categories: \emph{Time Lag}, \emph{Version Lag}, \emph{Package Lag}, \emph{Vulnerability and Bug Lag}, \emph{Opportunity Lag}, and \emph{Conceptual}. These categories provide a comprehensive framework for understanding and measuring TL from various perspectives and allow researchers to focus on specific categories to address potential gaps. One such gap identified is \emph{Version Lag}, which lacks the ability to differentiate between packages that are actively maintained, abandoned, or experiencing update delays relative to their upstream sources.

If a project is diligently maintained, its Version Lag should ideally decrease over time as maintainers update dependencies to their latest versions~\cite{panterPVACPackageVersion2025}. However, a hidden factor that can give a false sense of security is project abandonment, when maintainers stop actively maintaining a project. In such cases, Version Lag may no longer accurately reflect the project's health. A project may appear to have low Version Lag simply because the dependencies are no longer being updated. New metrics are needed to account for project abandonment, so stakeholders can better assess a software project's true health.

In prior work~\cite{panterPVACPackageVersion2025}, we implemented a Platform Version Activity Categorizer (PVAC) that introduced two new metrics to categorize the TL of an ecosystem and each package in isolation. At the ecosystem level, the Version Number Delta (VND) component computes a single TL score for all installed packages, providing a broad overview of system health. At the individual package level, the Activity Categorizer (AC) labels each package as \emph{Very Active}, \emph{Moderately Active}, \emph{Lightly Active}, or \emph{Sedentary} using semantic versioning data. The first three categories indicate varying levels of active maintenance, while the \emph{Sedentary} label signifies that a package is either abandoned, has very low maintenance activity (e.g., feature complete), or is experiencing an \emph{update delay}~\cite{legay_quantitative_2021} from upstream sources.

This paper introduces a conceptual distinction between resolvable technical lag, caused by delayed updates during active maintenance, and terminal technical lag, caused by upstream abandonment in which no update path exists.

\subsection{Research Questions}\label{sec:questions}

To advance the state of TL research and address the identified gap, we propose the following research questions:

\begin{description}
    \item[\textbf{RQ1:}]\RQone
    \item[\textbf{RQ2:}]\RQtwo
    \item[\textbf{RQ3:}]\RQthree
\end{description}

\subsection{Contributions}\label{sec:contributions}

This work introduces new TL methods and novel metrics to address a key limitation identified in prior TL research~\cite{panterTechnicalLagLatent2026a}. The contributions of this paper are as follows:

\begin{itemize}
    \item A new method and novel metrics for measuring TL that account for project abandonment and update delays.
    \item A new dataset constructed from operating system package repositories and their upstream sources.
    \item Empirical evaluation of the effectiveness of the metrics in capturing the true state of TL.
\end{itemize}

The remainder of this paper is structured as follows: Section~\ref{sec:background} provides background and motivation for our study. Section~\ref{sec:related} reviews related work on TL and project abandonment. Section~\ref{sec:methodology} outlines our proposed methodology for measuring abandonment-aware TL and our dataset construction. Our findings from applying the new metrics to our dataset are presented in Section~\ref{sec:results}. We discuss our results and their implications in Section~\ref{sec:discussion}. Finally, Section~\ref{sec:threats} presents the threats to validity, Section~\ref{sec:conclusion} concludes the paper, and discusses future research directions.

\section{Background and Motivation}\label{sec:background}

Modern applications are built using software components from third-party libraries and frameworks. These components are often distributed via a \emph{package manager}, which automates the installation, updating, and management of \emph{packages}. When an application depends on packages from multiple ecosystems, managing dependencies can become complex. To frame our discussion, we will review \emph{package managers} and \emph{packages}, and connect them to the concept of TL as it relates to \emph{upstream} and \emph{downstream} sources.

\subsection{Package Manager}\label{sec:package-manager}

The term \emph{package manager} is used broadly in the software development community to refer to tools that facilitate the installation and management of software packages. Broadly speaking, package managers can be categorized into two main types: language-specific (LSPM) and language-agnostic (LAPM)~\cite{muhammadTaxonomyPackageManagement2019}. LSPMs are designed to manage packages for a specific programming language or ecosystem. Examples include \texttt{npm} for JavaScript, \texttt{pip} for Python, and \texttt{Maven} for Java. These package managers typically provide a centralized repository of packages that can be easily searched and installed using command-line tools or graphical user interfaces. LAPMs, on the other hand, are designed to manage packages across multiple programming languages and ecosystems. Examples include \texttt{apt} for Debian-based Linux distributions, \texttt{Homebrew} for macOS, and \texttt{Chocolatey} for Windows.

\subsection{Package}\label{sec:package}

A \emph{package} is a collection of files and metadata that are bundled together to provide specific functionality or features. The contents of these packages can be source code, compiled binaries, configuration files, or other resources needed for the application to function correctly, and are largely dependent on the programming language and ecosystem in which the application is developed. For example, JavaScript packages typically contain JavaScript source code, while C++ packages may contain compiled binaries and header files. Additionally, a hidden aspect of packages is that they may not be entirely self-contained and may rely on external dependencies or libraries to function correctly. These dependencies can introduce additional complexity and potential TL if they are not kept up to date.

\subsection{Technical Lag and Abandonment}\label{sec:technical-lag}

\begin{figure}[htb]
    \centering
    \includegraphics[width=0.45\textwidth]{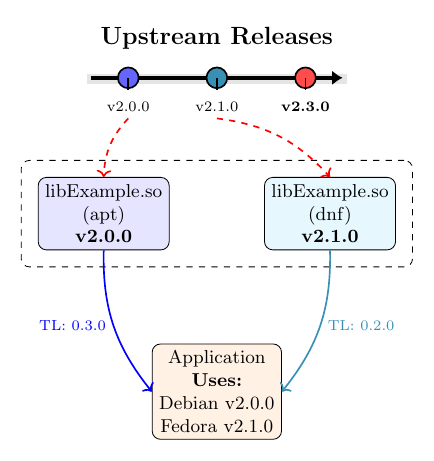}
    \caption{Version Lag: The upstream (latest) version for \textbf{libExample.so} is (v2.3.0) while the packages for two distributions (downstream) are still at v2.0.0 (apt) and v2.1.0 (dnf). The TL of the application is 0.3.0 for Debian and 0.2.0 for Fedora.}
    \label{fig:tech-lag-diagram}
\end{figure}

The general lifecycle of a package typically involves several stages, including development, release, maintenance, and eventual deprecation or abandonment~\cite{nguyenLifeDeathSoftware2012}. During development, the package is created and tested by the developer or development team. Once the package is deemed stable and ready for use, it is released to a package repository or registry, where users can download and install it. A critical aspect of LAPMs such as \texttt{apt} or \texttt{dnf} is that packages are often \emph{wrapped} versions of upstream packages. Fig.~\ref{fig:tech-lag-diagram} illustrates how an application can experience TL when installed on different distributions. This version may not be the latest available upstream, as distribution maintainers may delay updates to ensure stability and compatibility with other packages in the distribution~\cite{legay_quantitative_2021}.

When an upstream package is no longer actively maintained or updated, it is considered \emph{abandoned}~\cite{avelinoAbandonmentSurvivalOpen2019}. \textit{Time Lag} and \textit{Package Lag} metrics, which measure a package's release date relative to the application's release date, will indicate increasing TL for abandoned packages. Additionally, \textit{Vulnerability and Bug Lag} may increase as a package becomes increasingly out of date. \textit{Version Lag}, however, is measured by comparing the version of the upstream package to the version of the downstream package and is the most vulnerable to software abandonment issues, as the package will appear to have a TL of 0 if the downstream package is at the same version as the upstream package and no new versions will ever be released.

\subsection{Motivation}\label{sec:motivation}

Our motivation for this work stems from the need to better understand the state of packages that appear to be \textit{Sedentary}~\cite{panterPVACPackageVersion2025} and how TL and abandonment manifest in real-world package ecosystems. Differentiating between packages that are actively maintained, those experiencing maintenance decline, and those abandoned is crucial for developers and organizations to make informed decisions about their software dependencies. Not all software that appears inactive is truly abandoned; some may be stable or ``feature complete'' and require minimal updates.
\section{Related Work}\label{sec:related}

There has been extensive related work on project abandonment and technical lag (TL) in software development. Section~\ref{sec:related-abandonment} reviews the literature on project abandonment, while Section~\ref{sec:related-tl} focuses on TL.

\subsection{Project Abandonment}\label{sec:related-abandonment}

Xia et al.~\cite{xiaUnderstandingArchivedProjects2023} surveyed open source developers of projects that had been archived and found the two most common reasons for archiving a project are being superseded by alternatives and evolving into a successor. They found that projects tended to follow similar stages of evolution, active development lasting around 2 years, followed by a decline in activity and eventual archiving lasting over a year. The bus factor (or truck factor), is a significant predictor of project abandonment, with projects having a low bus factor being more vulnerable to abandonment which has been confirmed by other studies.~\cite{torchianoMyProjectsTruck2011, avelinoNovelApproachEstimating2016, ferreiraAlgorithmsEstimatingTruck2019}. Archived projects tend to have fewer core contributors, higher PR acceptance, and longer issue response times. A key finding from Avelino et al.~\cite{avelinoAbandonmentSurvivalOpen2019} was that 41\% of successful projects consistently attract new core developers to replace those that leave.

Miller et al.~\cite{millerDesigningAbandabotWhen2026} conducted a needs-finding interview with 22 open-source maintainers to understand what makes the abandonment of dependencies impactful. They built Abandabot-Predict a LLM-based classifier to help predict the impact of a dependency's abandonment in a given context evaluating it with 124 open source maintainers. They found that their classifier was effective at predicting developer judgments of abandonment impact, achieving Macro-F1 scores of 0.682 without context and 0.840 with context. Additional work by Miller et al.~\cite{millerUnderstandingResponseOpensource} found that there is little understanding of the prevalence of abandonment of widely-used packages. They studied the \textit{npm} ecosystem and identified that 15\% of their dataset became abandoned within a six-year observation period.

System complexity or a project's internal properties can also contribute to abandonment. For example, projects with a large codebase, many dependencies, or complex architecture may be more difficult to maintain and thus more likely to be abandoned. Grigorio et al.~\cite{grigorioSystemsProjectAbandonment2014} studied the Lines of Code (LOC) and the cyclomatic complexity of 10 projects and noticed a relation between uncontrolled complexity and software abandonment.

\subsection{Technical Lag}\label{sec:related-tl}

TL is a broad term that formalizes the concept of a system degrading over time as its dependencies become outdated. TL is defined as a comparison between a ``gold standard'' and the current state and an associated \textbf{lag function} to compute the lag between the two~\cite{gonzalez-barahonaTechnicalLagSoftware2017}. This definition captures how software can degrade over time. Upstream packages continue to evolve and release new versions, while downstream systems remain on older versions.

Panter and Eisty~\cite{panterTechnicalLagLatent2026a} conducted a Rapid Review of the current literature on TL and consolidate all the lag functions that have be conceptualized into six key categories: Time~\cite{decanOutdatednessWorkflowsGitHub2023, salzaThirdpartyLibrariesMobile2020, stringerTechnicalLagDependencies2020,legay_quantitative_2021, zeroualiAnalyzingTechnicalLag2018, heAutomatingDependencyUpdates2023, decanEvolutionTechnicalLag2018, zeroualiImpactOutdatedVulnerable2019, opdebeeckDockerHubImage2023}, Version~\cite{panterPVACPackageVersion2025 ,stringerTechnicalLagDependencies2020, zeroualiEmpiricalAnalysisTechnical2018, zeroualiImpactOutdatedVulnerable2019}, Package, Vulnerability and Bug~\cite{zeroualiImpactOutdatedVulnerable2019, huEmpiricalAnalysisVulnerabilities2024}, Opportunity~\cite{decanOutdatednessWorkflowsGitHub2023} and Conceptual~\cite{gonzalez-barahonaTechnicalLagSoftware2017, zeroualiFormalFrameworkMeasuring2019, gonzalez-barahonaCharacterizingOutdatenessTechnical2020}. Each category measures the TL of an application or ecosystem against a different ``gold standard'' and provides a different perspective on the system's TL.

Zerouali et al.~\cite{zeroualiMultidimensionalAnalysisTechnical2021} explored using multiple lag functions together, conducting a multidimensional analysis of TL in Debian-based Docker images. They found that using multiple lag functions is complementary and captures different aspects of TL. They found that all images suffer from some form of TL, even the lightest official images with the fewest number of dependencies.

\section{Methodology}\label{sec:methodology}

To address the limitations of existing TL metrics in the presence of declining or absent maintenance, we introduce \textbf{M}aintenance-\textbf{A}ware \textbf{L}ag and \textbf{T}echnical \textbf{A}bandonment (MALTA), a repository-derived scoring framework that estimates the likelihood of sustained project maintenance by integrating additional signals for TL analysis. MALTA consists of four main components:

\begin{itemize}
    \item \textbf{Development Activity Score (DAS)}: captures changes in commit velocity and recency of activity (Section~\ref{sec:das})
    \item \textbf{Maintainer Responsiveness Score (MRS)}: measures the degree of maintainer engagement with external contributions (Section~\ref{sec:mrs})
    \item \textbf{Repository Metadata Viability Score (RMVS)}: assesses repository visibility and explicit lifecycle status (Section~\ref{sec:rmvs})
    \item \textbf{Aggregation Score (MALTA)}: combines the DAS, MRS, and RMVS components into a final TL score (Section~\ref{sec:aggregation-score})
\end{itemize}

MALTA treats abandonment not as a binary state but as a latent, continuous property inferred from observable upstream signals. This design choice enables consistent application across heterogeneous ecosystems while avoiding assumptions about release policies or governance structures.

MALTA produces a normalized maintenance activity score that captures both historical decline and recent inactivity. This score is subsequently used to contextualize TL measurements, distinguishing lag caused by delayed maintenance from lag arising under active stewardship.

\subsection{Problem Definition}

Given a project $p$ hosted on a public forge (e.g., GitHub), we compute a continuous maintenance activity score intended to correlate with sustained stewardship. Let $S_{final}(p) \in [0,1]$ denote the final score, where higher values indicate stronger evidence of ongoing maintenance. We emphasize that $S_{final}$ is an estimate of maintenance likelihood and should not be interpreted as a definitive abandonment label.

\subsection{Observation Windows and Data Sources}

For each project $p$, we define two non-overlapping time windows:

\begin{itemize}
    \item $W_b = [t_0, t_1)$ (baseline window)
    \item $W_e = [t_1, t_2)$ (evaluation window)
\end{itemize}

with $W_b \cap W_e = \emptyset$. The baseline window precedes the evaluation window, ensuring that historical activity is measured independently of the assessment period. The baseline window is set to 24 months to match Debian's 24-month policy for long-term support (LTS), and the evaluation window is set to 18 months to provide a small buffer between LTS releases. This window of just under 4 years is also aligned with previous work finding that most projects fail within this timeframe~\cite{aitEmpiricalStudySurvival2022}. We collect three classes of upstream signals from the project's GitHub repository over these windows:

\begin{itemize}
    \item \textbf{Version-control activity}: commit timestamps and commit diffs
    \item \textbf{Contribution}: pull request creation, closure, or merge events, and unresolved pull requests
    \item \textbf{Repository metadata}: stars, forks, watchers, open issue counts, and archival status
\end{itemize}

\subsection{Development Activity Score (DAS)}\label{sec:das}

The DAS captures changes in commit velocity relative to historical norms and the recency of substantive activity. We construct the DAS using commit data from $W_b$ and $W_e$. Let $C_b$ and $C_e$ denote the number of non-trivial commits in $W_b$ and $W_e$, respectively. Non-trivial commits exclude changes that do not reflect substantive maintenance (e.g., documentation-only changes, merge commits). Let $t_{last}$ denote the number of days between the end of $W_e$ and the most recent non-trivial commit, and let $\tau = 180$ be a decay constant.

Commit rates are defined as:

\[
\lambda_b = \frac{C_b}{|W_b|}, \quad
\lambda_e = \frac{C_e}{|W_e|}
\]

The velocity decay ratio is:

\[
    D_c =
    \begin{cases}
    \frac{\lambda_e}{\lambda_b}, & \lambda_b > 0 \\
    1, & \lambda_b = 0 \land \lambda_e > 0 \\
    0, & \lambda_b = 0 \land \lambda_e = 0
    \end{cases}
\]

 We define an exponential recency term:

\[
R_c = e^{-t_{last}/\tau}
\]

The development activity score is:

\[
S_{dev} = \min(1, D_c) \cdot R_c
\]

We set the decay constant $\tau$ to 180 days to align with empirical findings from previous studies~\cite{xiaUnderstandingArchivedProjects2023, coelhoIdentifyingUnmaintainedProjects2018, decanEvolutionTechnicalLag2018, legay_quantitative_2021}, which consistently show that inactivity and responsiveness patterns become strongly predictive of abandonment only after sustained periods on the order of several months rather than weeks.

\subsection{Maintainer Responsiveness Score (MRS)}\label{sec:mrs}

The MRS measures the degree to which maintainers actively govern external contributions. Let $P$ be the set of pull requests opened during $W_e$, partitioned into:

\begin{itemize}
    \item $P_{term}$: pull requests opened during $W_e$ that were merged or closed during $W_e$
    \item $P_{open}$: pull requests opened during $W_e$ that remain open at $t_2$
\end{itemize}

\paragraph{Decision Responsiveness} The fraction of pull requests that reached a terminal state:

\[
    R_{dec} = \frac{|P_{term}|}{|P|}
\]

\paragraph{Decision Timeliness} For each $p \in P_{term}$, let $\Delta t(p)$ be the number of days between PR creation and closure or merge with $T_{ref} = 180$ days.

\[
    D_{dec} =
    \operatorname{median}\left(
    \min\left(1, \frac{\Delta t(p)}{T_{ref}}\right)
    \;\middle|\; p \in P_{term}
    \right)
\]

\paragraph{Open-PR Staleness Penalty} The Open-PR Staleness Penalty down-weights maintainer responsiveness scores even when some responses occur, ensuring that superficial or sporadic engagement does not mask prolonged neglect of external contributions. This prevents projects with a few timely responses but persistently stale pull requests from being misclassified as healthy.

\[
    P_{stale} =
    \operatorname{median}\left(
    \min\left(1, \frac{age(p)}{T_{ref}}\right)
    \;\middle|\; p \in P_{open}
    \right)
\]

\paragraph{Responsiveness Aggregation} If $|P| = 0$, $S_{resp}$ is treated as undefined rather than zero. If $|P| > 0$ but $|P_{term}| = 0$, then $S_{resp} = 0$.

\[
    S_{resp} = R_{dec} \cdot (1 - D_{dec}) \cdot (1 - P_{stale})
\]

\subsection{Repository Metadata Viability Score (RMVS)}\label{sec:rmvs}

RMVS captures lightweight structural signals of project visibility and explicit lifecycle status. Because metadata primarily reflects \emph{attention} rather than maintenance intent, we treat it as a weak signal and apply (i) log-saturation to reduce outlier sensitivity and (ii) a small overall weight in the final aggregation.

\paragraph{Log-saturated normalization} Let $x$ denote a non-negative repository-level count (e.g., stars, forks, watchers, open issues). For each metadata dimension $x$, we compute a dataset-level saturation constant $K_x$ as the empirical 95th percentile:

\[
K_x = Q_{0.95}(x)
\]

where $Q_{0.95}(\cdot)$ is computed over all repositories in our dataset. We then map each raw value $x_i$ to $[0,1]$ using a log-saturated transform $\phi(\cdot)$:

\[
\phi(x_i;K_x) = \min\!\left(1,\frac{\log(1+x_i)}{\log(1+K_x)}\right)
\]

This ensures that values at or above the 95th percentile saturate at 1 while preserving rank information among smaller repositories.

\paragraph{Metadata components} We define normalized popularity/attention signals:

\begin{align}
S^* &= \nonumber \phi(\textit{stars};K_{\textit{stars}}) \\
F^* &= \nonumber \phi(\textit{forks};K_{\textit{forks}}) \\
W^* &=  \nonumber\phi(\textit{watchers};K_{\textit{watchers}})
\end{align}

Open issues are treated as backlog pressure (a negative indicator). Let $I^* = \phi(\textit{open\_issues};K_{\textit{open\_issues}})$ and define an issue penalty as:

\[
I_{\text{pen}} = 1 - I^*
\]

\paragraph{Archival penalty} Let $A \in \{0,1\}$ indicate whether the repository is archived. We apply an explicit archival down-weight:

\[
A_{\text{pen}} = 1 - \alpha A
\]

with $\alpha = 0.7$, so archived repositories receive a multiplicative factor of $0.3$. The archival penalty parameter $\alpha$ is policy-driven rather than learned. We set $\alpha=0.7$ to reflect GitHub's strong but non-terminal archival designation. Sensitivity analysis in Section~\ref{sec:sensitivity} shows that results are stable for $\alpha \in [0.6,0.8]$.

\paragraph{RMVS aggregation} The metadata viability score is:

\[
S_{\text{meta}} = A_{\text{pen}}
\left(\beta_s S^* + \beta_f F^* + \beta_w W^* + \beta_i I_{\text{pen}}\right)
\]

where $\beta_s,\beta_f,\beta_w,\beta_i \ge 0$ and $\beta_s+\beta_f+\beta_w+\beta_i=1$. In our experiments, we set $\beta_s=\beta_f=\beta_w=\beta_i=0.25$ a priori and confirm their settings in Section~\ref{sec:sensitivity}.

\subsection{Final Aggregation Score}\label{sec:aggregation-score}

We aggregate the three signals using a weighted linear model:

\[
S_{final} = w_{dev} S_{dev} + w_{resp} S_{resp} + w_{meta} S_{meta}
\]

with fixed weights $w_{dev} = 0.55$, $w_{resp} = 0.35$, and $w_{meta} = 0.10$. The weights reflect the relative importance of direct maintenance signals (development activity and responsiveness) versus indirect metadata signals. Sensitivity analysis in Section~\ref{sec:sensitivity} shows that results are stable for reasonable weight variations.

If $S_{resp}$ is undefined, we renormalize over observed signals:

\[
S_{final} =
\frac{w_{dev} S_{dev} + w_{meta} S_{meta}}{w_{dev} + w_{meta}}
\]

If the repository is archived and $S_{resp}$ is undefined, we set $S_{resp} = 0$.

For Interpretation:

\[
S^{(100)}_{final} = 100 \cdot S_{final}
\]

MALTA is intentionally not learned end-to-end. The component weights and parameters are fixed a priori to favor interpretability, cross-ecosystem portability, and reproducibility. While supervised learning could optimize predictive performance for a specific dataset, such approaches risk overfitting to ecosystem-specific characteristics and obscuring the contribution of individual maintenance signals. Our goal is not to maximize classification accuracy but to provide a transparent, extensible framework that can be consistently applied across heterogeneous package ecosystems.

\subsection{Final Score Interpretation}\label{desc:final-score-interpretation}
We propose the following interpretive scale for $S^{(100)}_{final}$:

\begin{description}[topsep=1em]
\item[\textbf{$80 \dashrightarrow 100$ (Sustained Maintenance)}] Strong evidence of ongoing maintenance activity and responsiveness.
\item[\textbf{$60 \dashrightarrow 79$ (Stable Maintenance)}] Maintenance signals remain active but show early signs of slowdown.
\item[\textbf{$40 \dashrightarrow 59$ (Declining Maintenance)}] Clear and sustained reduction in maintenance activity or responsiveness.
\item[\textbf{$20 \dashrightarrow 39$ (Probable Abandonment)}] Strong indicators of maintenance inactivity.
\item[\textbf{$0 \dashrightarrow 19$ (Effective Abandonment)}] Near complete absence of maintenance signals.

\end{description}

For the risk reclassification analysis in Sections~\ref{sec:results-rq2} and~\ref{sec:results-rq3}, we collapse these five levels into three MALTA risk categories: \textbf{Low Risk} ($S_{final} \geq 0.60$, comprising Sustained and Stable), \textbf{Medium Risk} ($0.40 \leq S_{final} < 0.60$, Declining), and \textbf{High Risk} ($S_{final} < 0.40$, Probable and Effective Abandonment).

\subsection{Data Collection}\label{sec:methodology-dataset}

\begin{figure}
    \begin{center}
        \includegraphics[width=0.85\columnwidth, height=0.5\textheight, keepaspectratio]{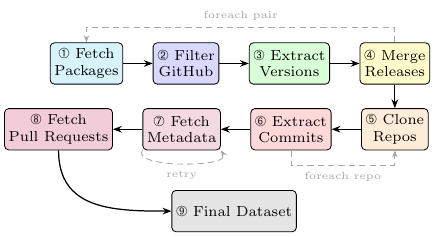}
        \caption{Construction pipeline of the empirical dataset with each step detailed in Section~\ref{sec:methodology-dataset} and summarized in Table~\ref{tab:dataset-stats}.}
        \label{fig:dataset-construction}
    \end{center}
\end{figure}

We constructed an empirical dataset\footnote{Available at \DataShare} to evaluate MALTA, following the process illustrated in Figure~\ref{fig:dataset-construction}. We began by fetching the list of all packages for the Debian Trixie and Bookworm releases\footnote{\url{https://ftp.debian.org/debian/dists/}} in step \ding{192}. The upstream source for each package was mapped to its corresponding GitHub repository in step \ding{193}, identifying those with GitHub repositories. In steps \ding{194} and \ding{195}, we extracted the version data for each package and mapped it to a tag in the corresponding GitHub repository. Next, in steps \ding{196} and \ding{197}, we cloned each repository and extracted commit data for each package. In step \ding{198}, we fetched repository metadata, and in step \ding{199}, we retrieved pull request data for each repository. The details of the final dataset, produced in step \ding{200}, are summarized in Table~\ref{tab:dataset-stats}. The dataset represents a diverse cross-section of open-source projects across two major Debian releases and any derivatives (such as Ubuntu), providing a robust foundation for evaluating MALTA's effectiveness in estimating maintenance activity and addressing software abandonment.

\begin{table}[htb]
    \caption{Dataset summary statistics after each data collection step.}
    \label{tab:dataset-stats}
    \begin{tabular}{p{1cm}p{4cm}p{2.5cm}}
    \toprule
    \footnotesize
    Step &Category & Value \\
    \midrule
    \ding{192} &Total Trixie Packages & 37,632 \\
    \ding{192} &Total Bookworm Packages & 34,340 \\
    \ding{193} &Filtered Trixie & 15,423 \\
    \ding{193} &Filtered Bookworm & 12,667 \\
    \ding{194},\ding{195} &Total Packages & \rqoneNPackages{} \\
    \ding{196},\ding{197} &Total commits & 1,721,811 \\
    \ding{198} &Total metadata & \rqoneNPackages{} \\
    \ding{199} &Total pull requests & 4,187,429 \\
    \midrule
    \multicolumn{3}{c}{\textbf{Stats}} \\
    \cmidrule(lr){1-3} 
    &Total Archived & \rqoneNArchived{} (\rqonePctArchived{}) \\
    &Median commits & 40 \\
    &Median stars & 132 \\
    &Median age (years) & 10.9 \\
    &Median PRs & 71 \\
    \bottomrule
    \end{tabular}
\end{table}

\section{Results}\label{sec:results}

In this section, we present the results of our empirical evaluation of MALTA across the three research questions. Section~\ref{sec:results-rq1} addresses RQ1 by analyzing the distribution of MALTA component scores and their ability to distinguish sustained maintenance from decline. Section~\ref{sec:results-rq2} investigates RQ2 by examining how Time Lag varies across maintenance states and whether packages with low Version Lag still exhibit signs of maintenance decline. Finally, Section~\ref{sec:results-rq3} evaluates RQ3 by comparing risk classifications between Version Lag and MALTA-informed models, including an analysis of discordant classifications where packages are rated as low risk by Version Lag but identified as high risk by MALTA. We also include a sensitivity analysis in Section~\ref{sec:sensitivity} to assess the robustness of our findings to parameter variations.

\subsection{\textbf{RQ1:} Distinguishing Sustained Maintenance from Decline}\label{sec:results-rq1}

We evaluated MALTA's effectiveness across \rqoneNPackages{} packages categorized by PVAC~\cite{panterPVACPackageVersion2025} activity states.

\subsubsection{Distribution of MALTA Component Scores}\label{sec:rq1-malta-distributions}

\begin{figure}[htbp]
    \centering
    \includegraphics[width=\linewidth]{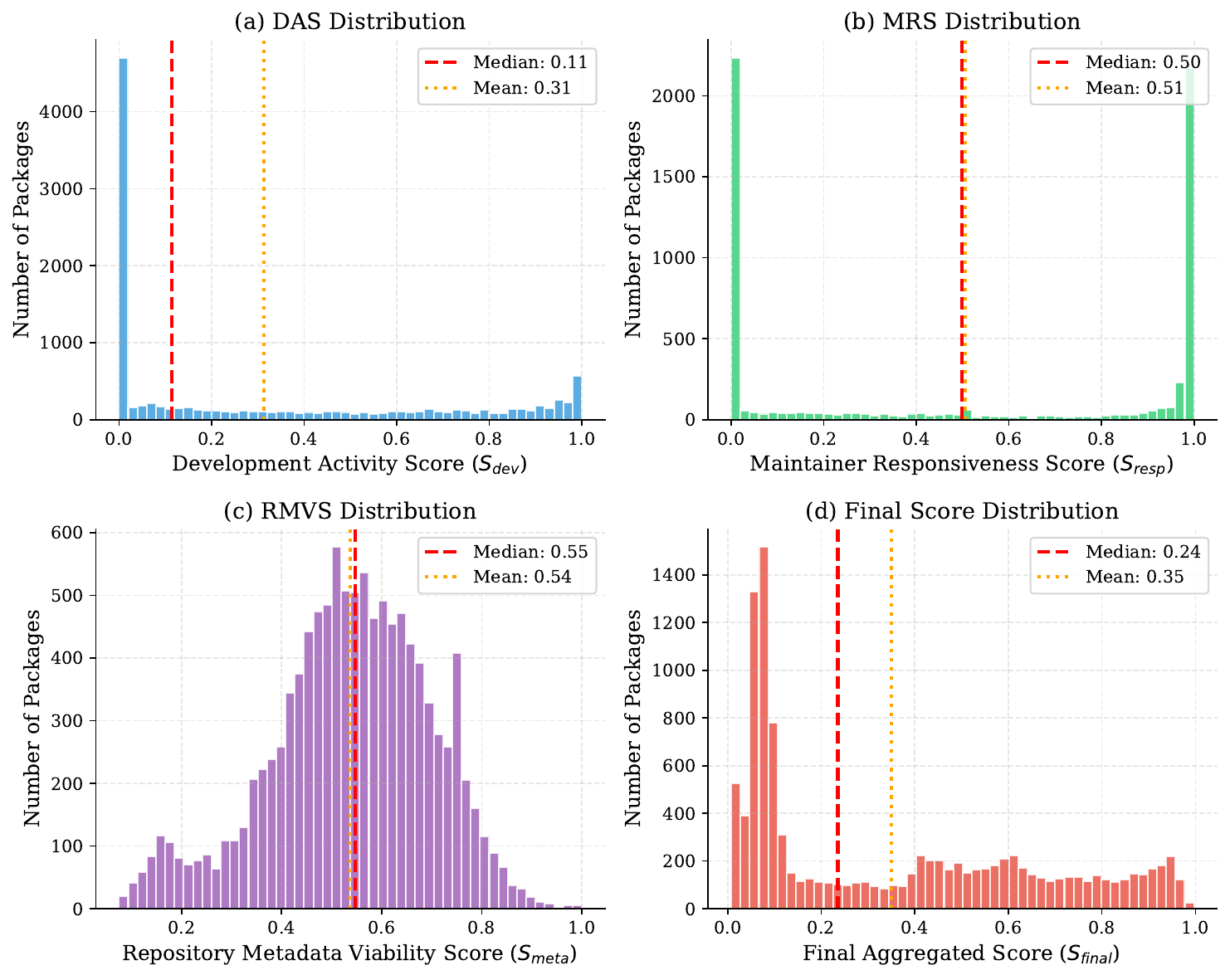}
    \caption{Distribution of MALTA Component Scores}
    \label{fig:malta-distributions}
\end{figure}

\begin{table}[t]
\caption{MALTA Component Score Statistics (n = 11,047)}
\label{tab:malta-statistics}
\centering
\begin{tabular}{lcccc}
\toprule
\textbf{Component} & \textbf{Mean} & \textbf{Median} & \textbf{Std} & \textbf{Range} \\
\midrule
DAS ($S_{dev}$) & 0.313 & 0.114 & 0.365 & [0.000, 1.000] \\
MRS ($S_{resp}$)\textsuperscript{$\dagger$} & 0.505 & 0.498 & 0.457 & [0.000, 1.000] \\
RMVS ($S_{meta}$) & 0.538 & 0.547 & 0.162 & [0.075, 1.000] \\
$S_{final}$ & 0.350 & 0.235 & 0.308 & [0.007, 0.997] \\
\bottomrule
\multicolumn{5}{l}{\parbox{\linewidth}{\textsuperscript{$\dagger$}\footnotesize Computed on $n=6{,}155$ packages with $|P|>0$; MRS is undefined for the remainder.}}
\end{tabular}
\end{table}

The three MALTA components exhibit markedly different distributional shapes (Table~\ref{tab:malta-statistics},  Figure~\ref{fig:malta-distributions}).

$S_{dev}$ (DAS) shows a strongly right-skewed distribution with a mean of \rqoneDASMean{} and a median of \rqoneDASMedian{}, indicating that the majority of packages exhibit low commit and release activity within the evaluation window. The large gap between mean and median, coupled with a standard deviation of 0.365, reveals a long tail of highly active projects that pull the mean upward while most repositories cluster near zero.

$S_{resp}$ (MRS) is defined only for the \rqoneNMRSDefined{} packages that received at least one pull request during the evaluation window; the remaining 44.3\% have undefined responsiveness scores and are excluded from the aggregation (Section~\ref{sec:aggregation-score}).  Among packages with defined MRS, the distribution is roughly symmetric  (mean 0.505, median \rqoneMRSMedian{}, std 0.457), spanning the full  $[0,1]$ range. This widespread reflects substantial heterogeneity in PR governance: some projects resolve contributions promptly, while others leave them unaddressed for extended periods. The high proportion of packages with undefined MRS highlights a structural limitation of PR-based signals in ecosystems where many projects receive no external contributions or manage them through channels outside GitHub.

$S_{meta}$ (RMVS) behaves differently from the activity-based components. Because stars, forks, and watchers accumulate over a project's lifetime rather than within a bounded evaluation window, the distribution is roughly symmetric (median \rqoneRMVSMedian{}, mean \rqoneRMVSMean{}) with a standard deviation of only 0.162. This low variance limits RMVS's discriminative value but also makes it a stable baseline signal less susceptible to short-term fluctuations.

The aggregated score $S_{final}$ inherits the right skew of its activity-weighted components, yielding a median of \rqoneFinalMedian{} and spanning nearly the full $[0, 1]$ range. The dominance of DAS in the weighting scheme means that $S_{final}$ primarily tracks development activity, with responsiveness and metadata acting as moderating signals.

\subsubsection{MALTA Scores by Activity State}\label{sec:rq1-malta-by-maintenance-state}

\begin{figure}[htbp]
    \centering
    \includegraphics[width=\linewidth]{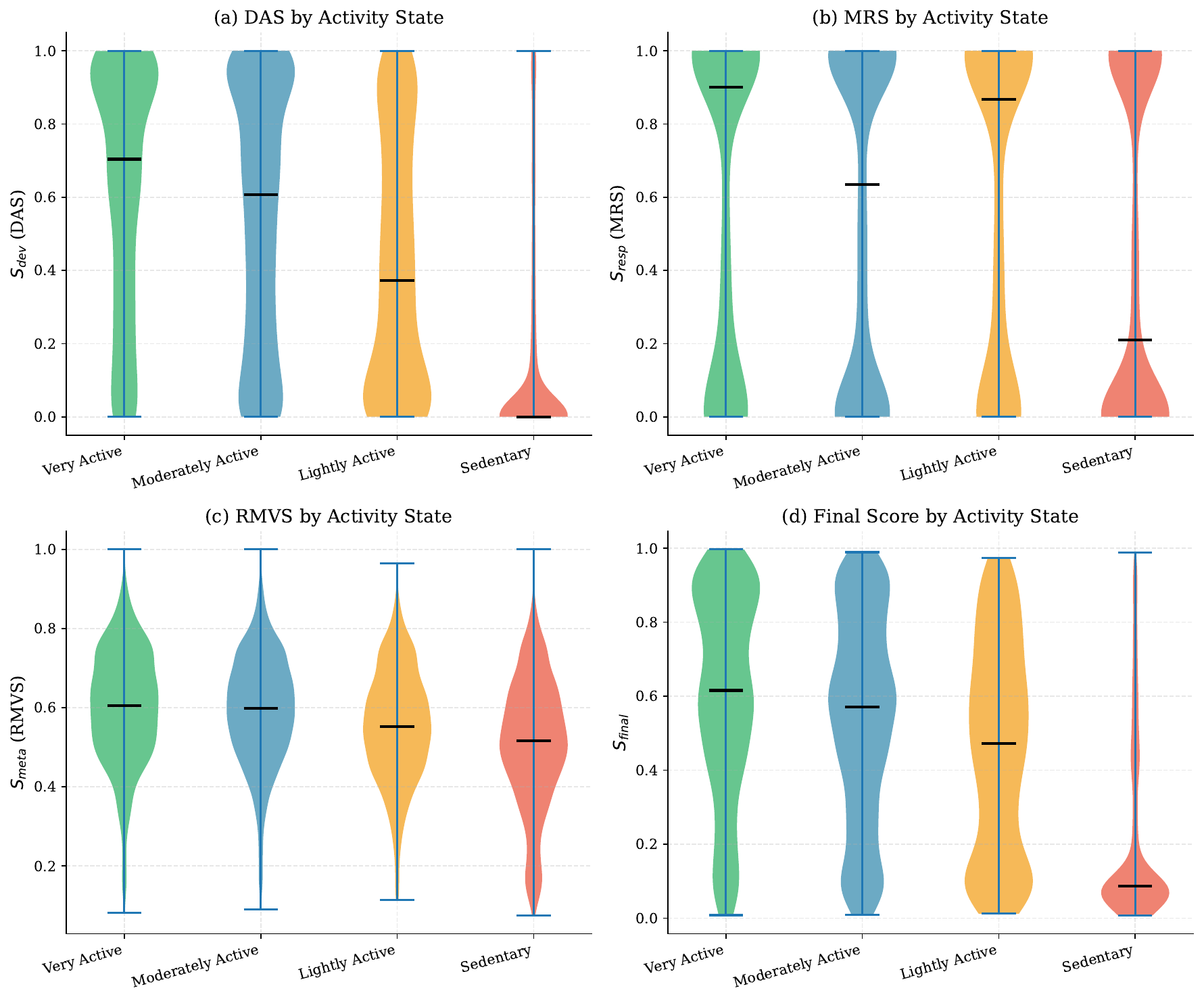}
    \caption{Distribution of MALTA Component Scores by PVAC Activity State}
    \label{fig:malta-by-pvac}
\end{figure}

\begin{table}[t]
\caption{Median MALTA Scores by PVAC Category}
\label{tab:malta-by-pvac}
\centering
\begin{tabular}{lrcccc}
\toprule
\textbf{PVAC State} & \textbf{n} & \textbf{DAS} & \textbf{MRS} & \textbf{RMVS} & \textbf{$S_{final}$} \\
\midrule
Very Active & 1,023 & 0.704 & 0.900 & 0.605 & 0.616 \\
Moderately Active & 2,284 & 0.606 & 0.635 & 0.598 & 0.571 \\
Lightly Active & 1,037 & 0.372 & 0.867 & 0.553 & 0.471 \\
Sedentary & 6,703 & 0.000 & 0.210 & 0.517 & 0.086 \\
\bottomrule
\end{tabular}
\end{table}

DAS exhibits the strongest discriminative gradient across PVAC activity states (Table~\ref{tab:malta-by-pvac}, Figure~\ref{fig:malta-by-pvac}). Very Active packages achieve a median $S_{dev}$ of \rqoneVeryActiveDASMedian{}, declining progressively through Moderately Active (0.606) and Lightly Active (0.372) to a floor of \rqoneSedentaryDASMedian{} for Sedentary packages. The collapse to zero at the Sedentary level reflects the DAS design: packages with no commits or releases within the evaluation window receive a score of exactly zero, producing the sharp separation visible in the violin plots.

MRS shows a less graduated pattern. While all three non-Sedentary categories retain moderate median values (ranging from 0.635 to 0.900),  Sedentary packages drop to a median of 0.210.  The wide spread among active categories suggests that PR responsiveness varies considerably within each activity level and is a noisier signal than commit frequency. The violin plots confirm this, with broad, overlapping distributions for the active states, contrasting with the concentration of low values in the Sedentary packages.

RMVS provides the least separation. The gap between Very Active (\rqoneVeryActiveRMVSMedian{}) and Sedentary (\rqoneSedentaryRMVSMedian{}) medians is less than 0.1, which is unsurprising given that metadata signals reflect cumulative visibility rather than current effort. This weak gradient justifies the 10\% weight assigned to RMVS in the final aggregation.

The aggregated $S_{final}$ closely mirrors the DAS gradient, with median scores of \rqoneVeryActiveFinalMedian{} for Very Active and \rqoneSedentaryFinalMedian{} for Sedentary, a nearly tenfold difference. The DAS-dominated weighting scheme ensures that the final score primarily captures the commit-level signal that provides the clearest activity-state separation, while MRS and RMVS contribute secondary modulation.

\subsubsection{Classification Performance}\label{sec:rq1-classification-performance}

We evaluated each MALTA component for two tasks: (1) Active (n=\rqoneNActive{}) vs.\ Declining (n=\rqoneNDeclining{}) classification, and (2) archived repository prediction (n=\rqoneNArchived{}, \rqonePctArchived{} of dataset).

\begin{figure}[htbp]
    \centering
    \includegraphics[width=\linewidth]{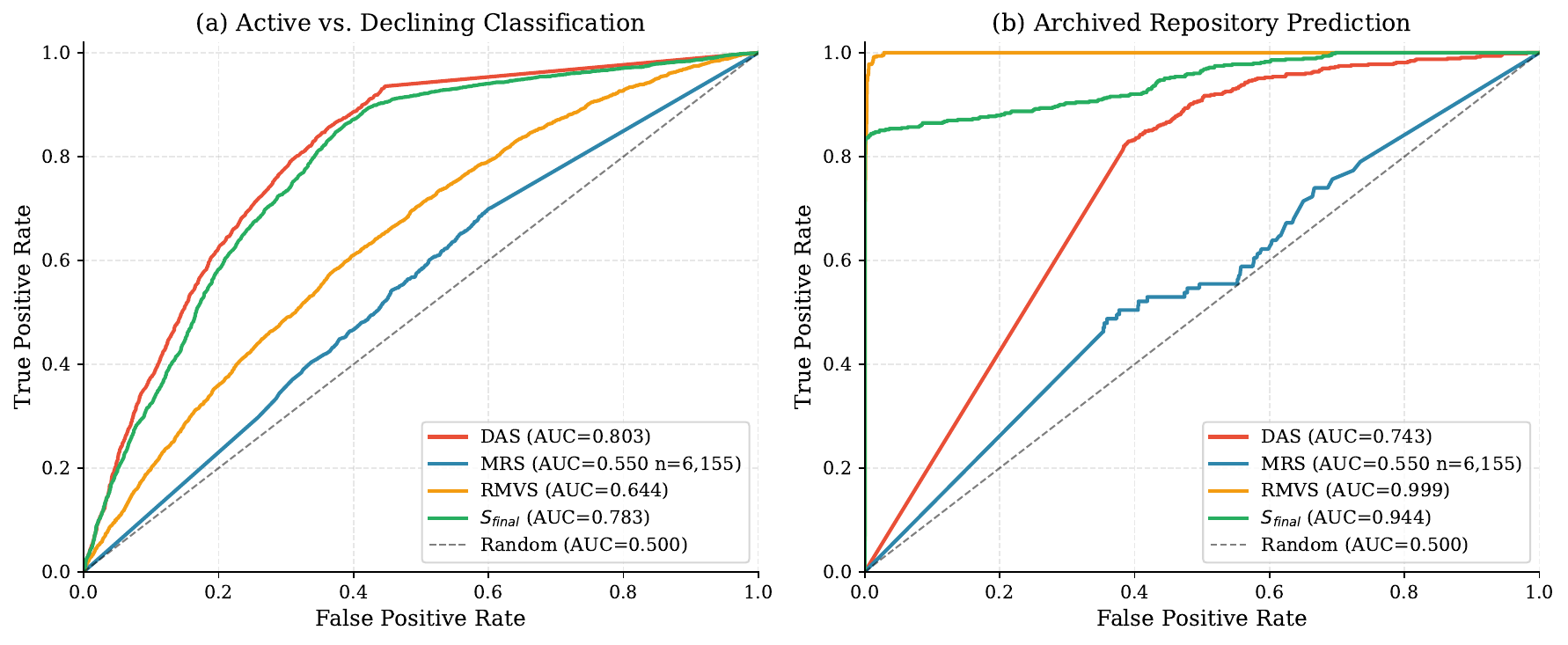}
    \caption{ROC Curves: (a) Active vs. Declining, (b) Archived prediction}
    \label{fig:classification}
\end{figure}

\begin{table}[t]
\caption{Classification Performance (AUC-ROC). All scores evaluated on $n=11,047$ packages except MRS ($n=6,155$, repos with $|P|>0$).}
\label{tab:classification}
\centering
\begin{tabular}{lcc}
\toprule
\textbf{Signal} & \textbf{Active vs. Declining} & \textbf{Archived Prediction} \\
\midrule
DAS & 0.803 & 0.743 \\
MRS & 0.550 & 0.550 \\
RMVS & 0.644 & 0.999 \\
$S_{final}$ & 0.783 & 0.944 \\
\midrule
Baseline (random) & 0.500 & 0.500 \\
\bottomrule
\end{tabular}
\end{table}

Table~\ref{tab:classification} and Figure~\ref{fig:classification} report the AUC-ROC for each MALTA component and the aggregated score on two binary classification tasks: (a)~distinguishing active packages (\rqoneNActive{} Very Active or Moderately Active) from declining ones (\rqoneNDeclining{} Lightly Active or Sedentary), and (b) predicting whether a repository has been explicitly archived (\rqoneNArchived{}, \rqonePctArchived{} of the dataset).

\paragraph{Active vs.\ Declining}

DAS is the single strongest predictor of ongoing maintenance, achieving an AUC of \rqoneDASAUCActive{}, substantially above both MRS (\rqoneMRSAUCActive{}) and RMVS (\rqoneRMVSAUCActive{}). This confirms that recent commits and release frequency capture the core dimension along which actively maintained packages diverge from declining ones. The aggregated $S_{final}$ attains an AUC of \rqoneFinalAUCActive{}, slightly below DAS alone, indicating that blending in the noisier MRS and weakly discriminative RMVS components dilutes rather than enhances the activity-state signal. MRS and RMVS perform comparably to each other (AUC~$\approx 0.64$), each providing modest discriminative power above the random baseline but falling well short of DAS.
This tradeoff is expected: commit activity best captures current development intensity, whereas the aggregated score is designed to capture broader lifecycle signals, including responsiveness and explicit archival, which become more informative for detecting lifecycle termination than for distinguishing fine-grained activity levels.

\paragraph{Archived Prediction}

The ranking of components reverses sharply for archived-repository detection. RMVS achieves a near-perfect AUC of \rqoneRMVSAUCArchived{}, reflecting the fact that RMVS directly encodes the archived flag via the $A_{\mathrm{pen}}$ penalty term. This result serves primarily as a validation check: RMVS is \emph{expected} to separate archived from non-archived repositories by construction. More informative is the performance of $S_{final}$, which reaches an AUC of \rqoneFinalAUCArchived{} despite DAS alone achieving only 0.743 on this task. The substantial lift from 0.743 (DAS) to \rqoneFinalAUCArchived{} ($S_{final}$) demonstrates that the multi-signal aggregation adds genuine value for archived-repository detection, where commit activity alone is an incomplete signal, some archived repositories retain non-trivial commit histories from before archival.

\subsubsection{Independent Validation Against Archived Status}\label{sec:rq1-archive-validation}

The PVAC-based evaluation in Section~\ref{sec:rq1-classification-performance} uses an activity categorization derived from versioning data, which shares conceptual overlap with MALTA's commit-based signals. To provide an independent validation, we evaluate MALTA's ability to identify repositories whose maintainers have \emph{explicitly} marked them as archived on GitHub a deliberate human decision that serves as a high-confidence indicator of maintenance cessation.

\paragraph{Removing the archival signal}

Because RMVS directly encodes the archived flag through the $A_{\mathrm{pen}}$ penalty (Section~\ref{sec:rmvs}), evaluating the full $S_{final}$ against archived status would be circular. We therefore define a modified score $S_{final}^{-A}$ that sets $A_{\mathrm{pen}} = 1$ for all packages, removing the only component that directly observes the archived flag:

\[
S_{final}^{-A} = w_{dev}\,S_{dev} \;+\; w_{resp}\,S_{resp} \;+\; w_{meta}\,S_{\mathrm{meta}}^{-A}
\]

\noindent where $S_{\mathrm{meta}}^{-A} = \beta_s S^* + \beta_f F^* + \beta_w W^* + \beta_i I_{\mathrm{pen}}$ omits the archival penalty. All other parameters remain at their default values. This score relies exclusively on commit velocity, PR responsiveness, and popularity metadata, none of which encode archived status.

\paragraph{Classification performance}

\begin{figure}[hbp]
    \centering
    \includegraphics[width=\linewidth]{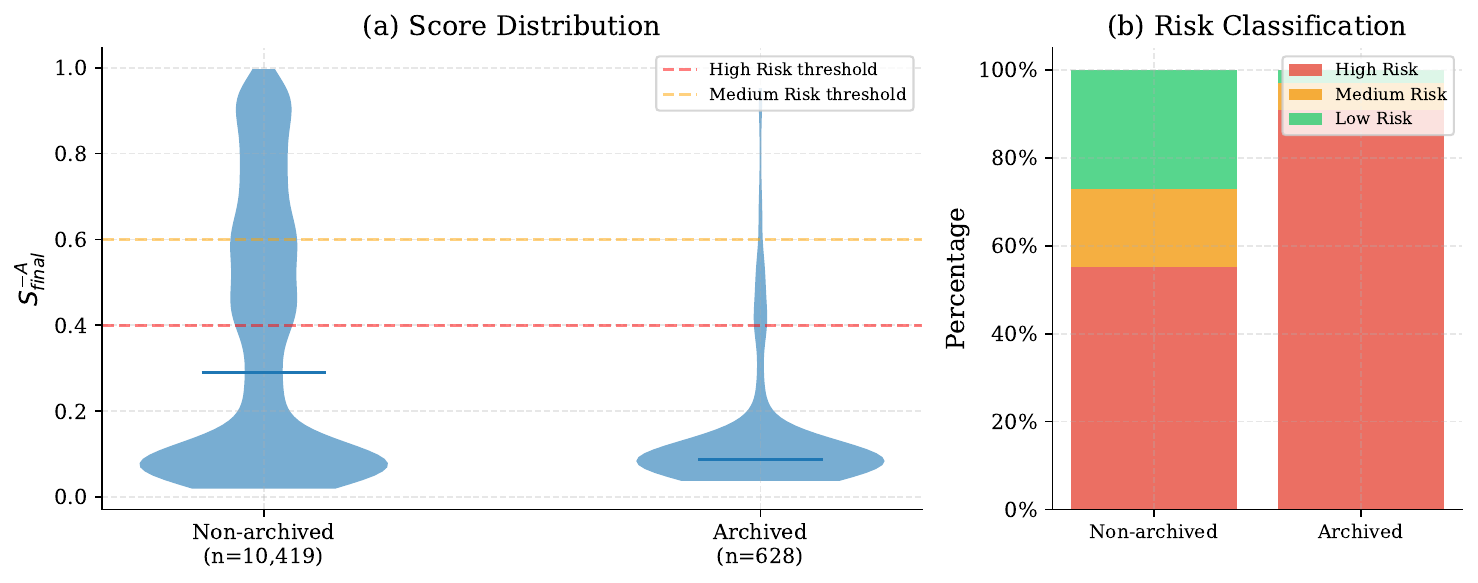}
    \caption{Distribution of $S_{final}^{-A}$ for archived vs.\ non-archived repositories. The modified score excludes the archival penalty from RMVS.}
    \label{fig:archive-validation}
\end{figure}

\begin{table*}[t]
\caption{Archived Flag Validation: AUC-ROC for archived-repository prediction with and without the archival penalty $A_{\mathrm{pen}}$, and risk classification under $S_{final}^{-A}$.}
\label{tab:archive-validation}
\centering
\begin{tabular}{lcccc}
\toprule
 & \textbf{DAS} & \textbf{MRS} & \textbf{RMVS} & \textbf{$S_{final}$} \\
\midrule
\multicolumn{5}{l}{\textit{AUC-ROC for Archived Prediction}} \\
With $A_{\mathrm{pen}}$ & 0.743 & 0.636 & 0.999 & 0.944 \\
Without $A_{\mathrm{pen}}$ & 0.743 & 0.636 & 0.500 & 0.686 \\
\midrule
\multicolumn{5}{l}{\textit{Risk Classification under $S_{final}^{-A}$}} \\
 & \multicolumn{2}{c}{\textbf{Archived} ($n=628$)} & \multicolumn{2}{c}{\textbf{Non-archived} ($n=10,419$)} \\
\cmidrule(lr){2-3} \cmidrule(lr){4-5}
High ($<0.40$) & \multicolumn{2}{c}{90.9\%} & \multicolumn{2}{c}{55.3\%} \\
Medium ($0.40$--$0.59$) & \multicolumn{2}{c}{6.2\%} & \multicolumn{2}{c}{17.8\%} \\
Low ($\geq 0.60$) & \multicolumn{2}{c}{2.9\%} & \multicolumn{2}{c}{27.0\%} \\
\midrule
\multicolumn{5}{l}{\textit{Effect Size} (Mann--Whitney $U=2,056,398$, $p<10^{-54}$, Cliff's $\delta=0.371$, medium)} \\
\bottomrule
\end{tabular}
\end{table*}

Table~\ref{tab:archive-validation} reports the AUC-ROC for each MALTA component and the modified aggregate $S_{final}^{-A}$ on the task of predicting archived status (\rqoneNArchived{} archived repositories, \rqonePctArchived{} of the dataset) and is visualized in Figure~\ref{fig:archive-validation}. DAS alone achieves an AUC of 0.743, confirming that commit inactivity is a meaningful but incomplete signal of archival. The modified aggregate $S_{final}^{-A}$ achieves an AUC of \rqoneAUCNoArchFinal{}, demonstrating that MALTA's multi-signal design captures archived-repository characteristics \emph{beyond} the trivial encoding of the archived flag.

\paragraph{Risk classification}
Applying the MALTA risk thresholds from Section~\ref{desc:final-score-interpretation} to $S_{final}^{-A}$, \rqoneArchHighRiskPct{}\% of archived repositories are classified as High Risk ($S_{final}^{-A} < 0.40$), compared to \rqoneNonArchHighRiskPct{}\% of non-archived repositories. Only \rqoneArchLowRiskPct{}\% of archived repositories receive a Low Risk classification. This demonstrates that MALTA's risk stratification aligns with explicit maintainer abandonment decisions even when the archived flag is withheld from the score computation.

\paragraph{Limitations of archival as ground truth}
The archived flag provides high-confidence positive labels but suffers from low recall: many abandoned repositories are never formally archived. Consequently, the non-archived population includes an unknown number of effectively abandoned projects, which means the reported AUC is a \emph{conservative} estimate of MALTA's true discriminative ability. Additionally, the archived class (\rqoneNArchived{} packages, \rqonePctArchived{}) is small relative to the full dataset, and results should be interpreted with this class imbalance in mind.

\begin{tcolorbox}[
    enhanced,attach boxed title to top center={yshift=-3mm,yshifttext=-1mm},
    title=Answer to RQ1,fonttitle=\bfseries,
    boxed title style={size=small}]
The Development Activity Score (DAS) is the most reliable signal for distinguishing sustained maintenance from long-term decline, achieving AUC = \rqoneDASAUCActive{} for the Active vs.\ Declining classification task. This finding holds under independent validation against explicitly archived repositories: the modified score $S_{final}^{-A}$ (with the archival flag withheld) achieves AUC = \rqoneAUCNoArchFinal{}, and \rqoneArchHighRiskPct{}\% of archived repositories are classified as High Risk by MALTA without access to the archived signal.
\end{tcolorbox}

\subsection{\textbf{RQ2:} Time Lag and Hidden Maintenance Decline}\label{sec:results-rq2}

We examined Time Lag (defined as days since last commit) across all \rqtwoNPackages{} packages to assess how temporal staleness tracks maintenance state, and whether packages with low Version Lag still exhibit signs of maintenance decline. For \rqtwoNCapped{} packages (\rqtwoCappedPct{}\%) lacking commit history before the evaluation date, we assigned a capped Time Lag of \rqtwoTimeLagCapDays{} days ($\approx$10 years), treating absent commit history as an extreme form of staleness.

\subsubsection{Time Lag by Maintenance State}\label{sec:rq2-timelag}

\begin{figure}[htbp]
    \centering
    \includegraphics[width=\linewidth]{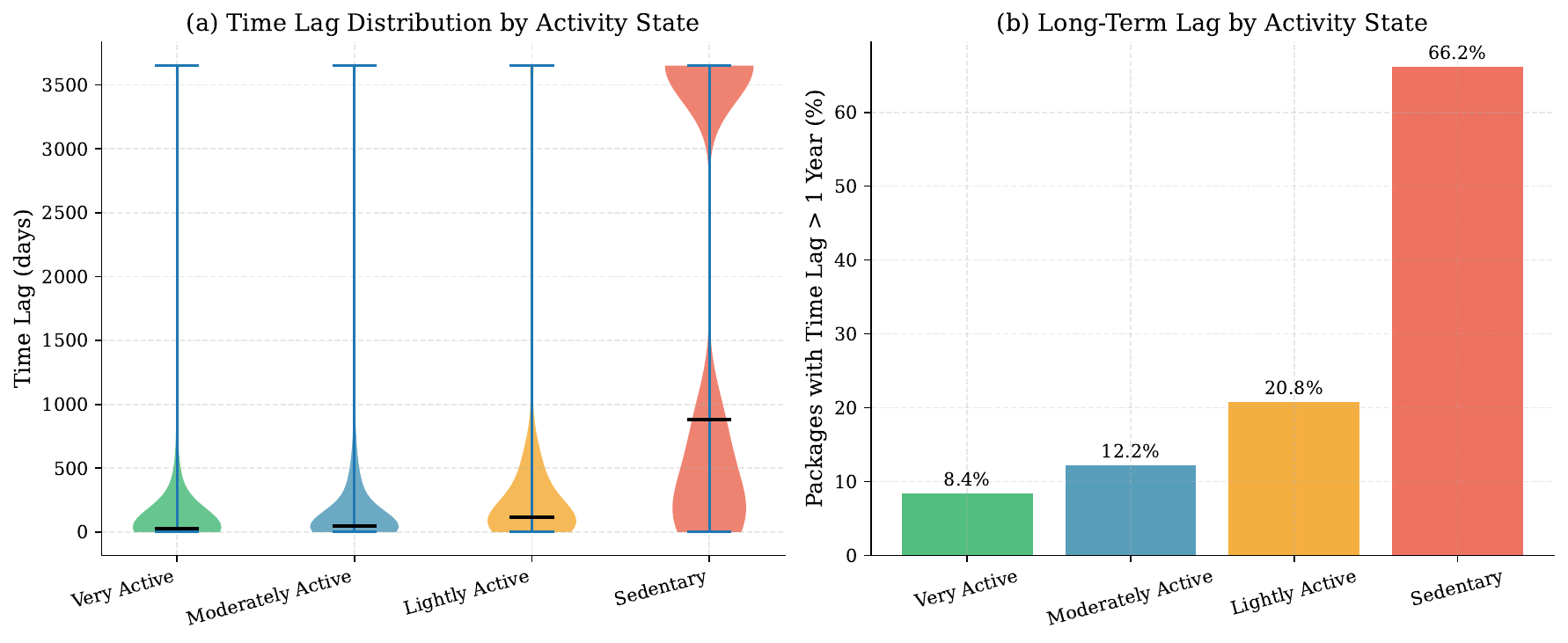}
    \caption{Time Lag by PVAC Activity State: (a) distribution, (b) long-term lag ($>$1 year) percentage.}
    \label{fig:rq2-timelag}
\end{figure}

\begin{table}[t]
\caption{Time Lag Statistics by PVAC Activity State (days)}
\label{tab:rq2-timelag-stats}
\centering
\begin{tabular}{lrcccc}
\toprule
\textbf{PVAC State} & \textbf{n} & \textbf{Median} & \textbf{Mean} & \textbf{IQR} & \textbf{$>$ 1 yr} \\
\midrule
Very Active & 1,023 & 25 & 170 & 6--99 & 8.4\% \\
Moderately Active & 2,284 & 45 & 198 & 9--170 & 12.2\% \\
Lightly Active & 1,037 & 115 & 282 & 29--323 & 20.8\% \\
Sedentary & 6,703 & 879 & 1809 & 188--3652 & 66.2\% \\

\bottomrule
\end{tabular}
\end{table}

Table~\ref{tab:rq2-timelag-stats} and Figure~\ref{fig:rq2-timelag} present the distribution of Time Lag stratified by PVAC activity state.

Time Lag increases monotonically with declining maintenance. Very Active packages exhibit a median Time Lag of only \rqtwoVATimeLagMedian{} days, rising to \rqtwoMATimeLagMedian{} days for Moderately Active, \rqtwoLATimeLagMedian{} days for Lightly Active, and \rqtwoSedTimeLagMedian{} days for Sedentary packages, a \rqtwoTimeLagMedianRatio{} increase from the most to least active category.

The long-term lag analysis in Figure~\ref{fig:rq2-timelag} confirms this finding. Only \rqtwoVALongPct{}\% of Very Active packages exceed one year without a commit, compared to \rqtwoSedLongPct{}\% of Sedentary packages, a \rqtwoLongTermRatio{} difference. The transition is not gradual: Moderately Active (\rqtwoMALongPct{}\%) and Lightly Active (\rqtwoLALongPct{}\%) occupy an intermediate range, while Sedentary packages show a qualitative shift, with the majority having stale repositories.

These results establish that Time Lag provides a meaningful temporal signal that tracks maintenance state. However, Time Lag alone cannot distinguish between a project that is version-current but unmaintained (e.g., a stable library with no recent commits) and one that is genuinely active.

\subsubsection{Low Version Lag with High Time Lag}\label{sec:rq2-hidden-decline}

\begin{figure}[htbp]
    \centering
    \includegraphics[width=\linewidth]{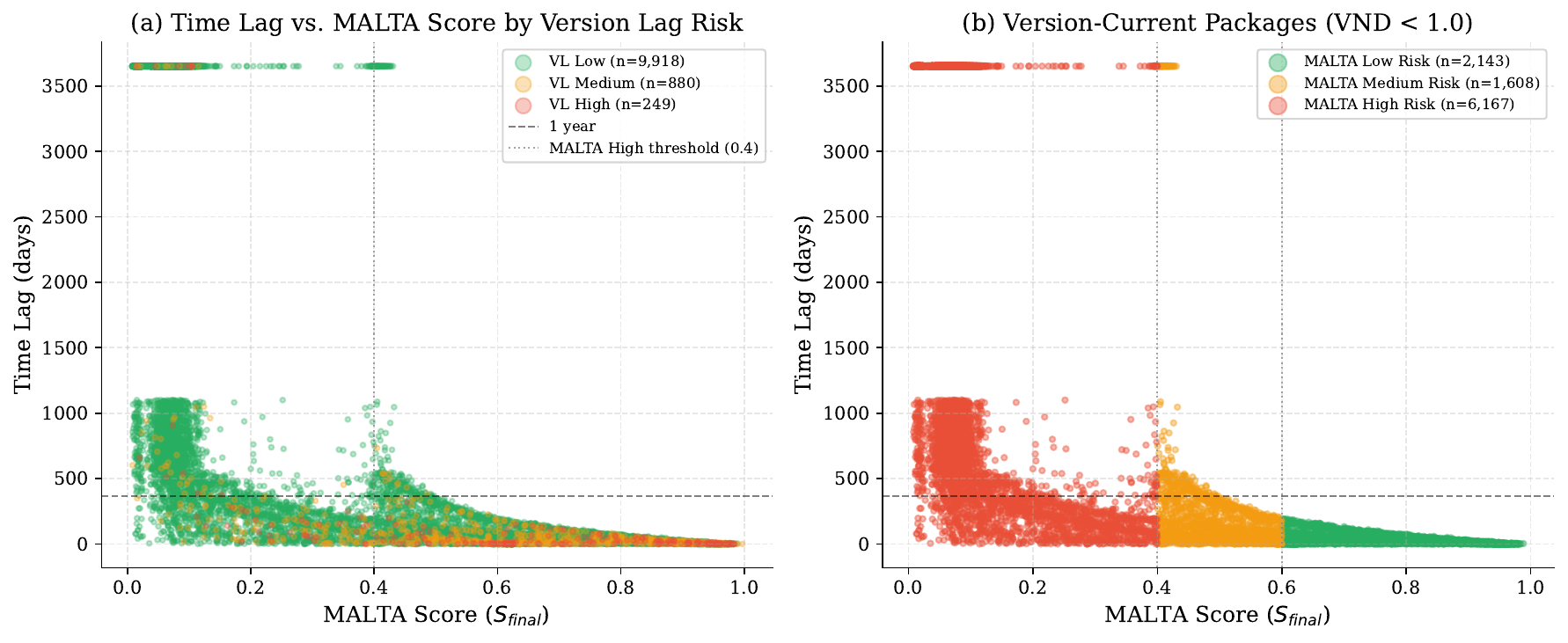}
    \caption{Time Lag vs.\ MALTA Score: (a) all packages colored by Version Lag risk, (b) version-current packages (VND $<$ \rqtwoLowVNDThreshold{}) colored by MALTA risk.}
    \label{fig:rq2-hidden-decline}
\end{figure}

\begin{table}[t]
\caption{Hidden Maintenance Decline Among Version-Current Packages (VND $<$ 1.0, n = 9,918)}
\label{tab:rq2-hidden-decline}
\centering
\begin{tabular}{lccccc}
\toprule
\textbf{MALTA Risk} & \textbf{n} & \textbf{\%} & \textbf{Median TL} & \textbf{$>$ 1 yr} & \textbf{Archived} \\
 & & & \textbf{(days)} & & \\
\midrule
Low & 2,143 & 21.6\% & 20 & 0.0\% & 0.0\% \\
Medium & 1,608 & 16.2\% & 104 & 13.4\% & 0.7\% \\
High & 6,167 & 62.2\% & 1067 & 76.8\% & 9.8\% \\
\bottomrule
\end{tabular}
\end{table}

To what extent does version currency mask maintenance decline? Among the \rqtwoLowVLN{} packages classified as Low Risk by Version Lag  (VND $< \rqtwoLowVNDThreshold{}$, Table~\ref{tab:rq2-hidden-decline}, Figure~\ref{fig:rq2-hidden-decline}), we assess how many exhibit deteriorating maintenance according to MALTA. Among the \rqtwoLowVLN{} packages classified as Low Risk by Version Lag (VND $< \rqtwoLowVNDThreshold{}$), we assess how many exhibit signs of deteriorating maintenance according to MALTA.

Figure~\ref{fig:rq2-hidden-decline}(a) plots Time Lag against MALTA score for all packages, colored by Version Lag risk category. A substantial cluster of Low Version Lag packages (green) appears, with high Time Lag and low MALTA score, indicating packages that are version-current yet show clear maintenance decline. This cluster is largely invisible to Version Lag alone.

Zooming in on Version-Current packages in Figure~\ref{fig:rq2-hidden-decline}(b) and Table~\ref{tab:rq2-hidden-decline}, \rqtwoLowVLHighMaltaPct{}\% (\rqtwoLowVLHighMaltaN{}) are classified as High Risk by MALTA ($S_{final} < 0.40$), with a median Time Lag of \rqtwoLowVLHighMaltaTL{} days and an archived rate of \rqtwoLowVLHighMaltaArchPct{}\%. An additional \rqtwoLowVLMedMaltaPct{}\% fall into MALTA Medium Risk ($0.40 \leq S_{final} < 0.60$). Together, these packages represent a \emph{hidden decline} population: they would be considered healthy by any Version Lag assessment, yet MALTA identifies substantial maintenance deterioration.

Only \rqtwoLowVLLowMaltaPct{}\% of version-current packages are confirmed as genuinely healthy by both metrics (VL Low and MALTA Low), with a median Time Lag of just \rqtwoLowVLLowMaltaTL{} days. The \rqtwoLowVLHiddenRatio{} gap between the MALTA-healthy and MALTA-high-risk medians underscores that version currency and maintenance health measure fundamentally different dimensions of software viability.

\subsubsection{MALTA Scores of Version-Current Packages}\label{sec:rq2-malta-version-current}

\begin{figure}[htbp]
    \centering
    \includegraphics[width=\linewidth]{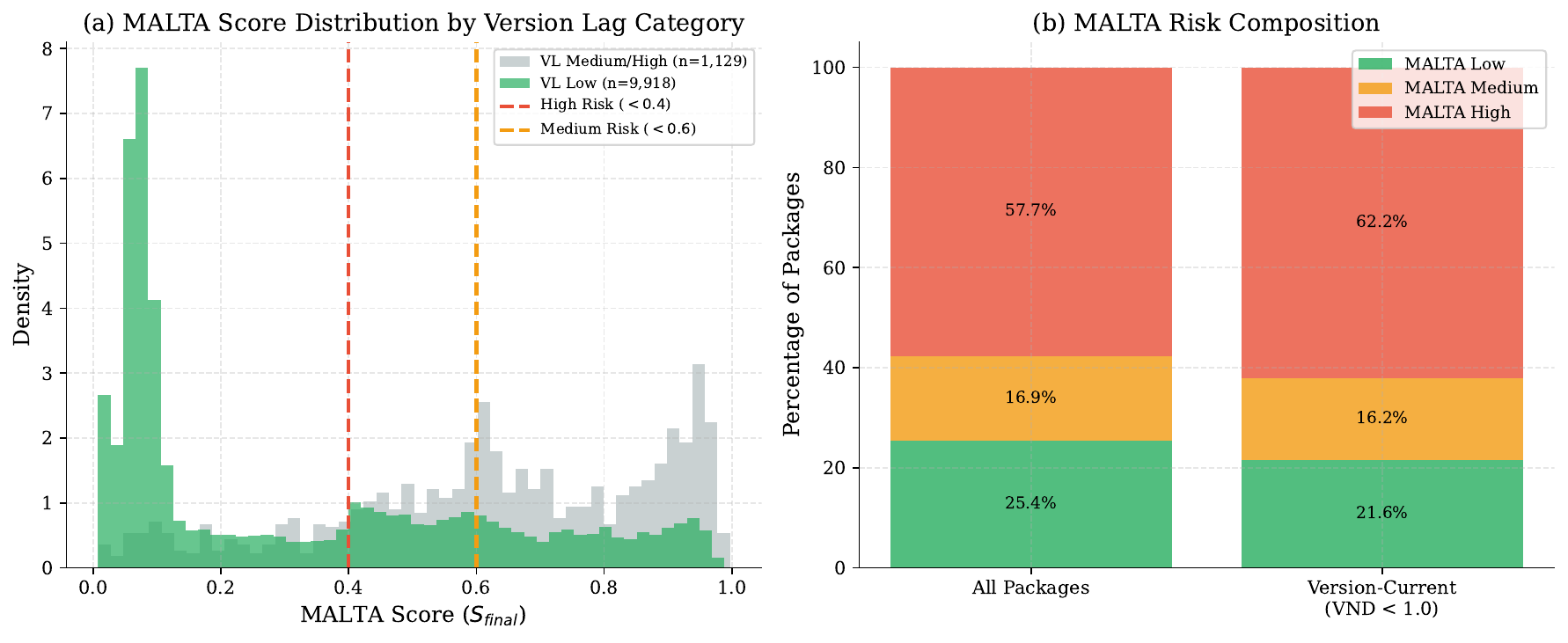}
    \caption{MALTA scores of version-current packages: (a) score distribution compared to the full dataset, (b) risk composition.}
    \label{fig:rq2-malta-low-vl}
\end{figure}

\begin{table}[t]
\caption{MALTA Maintenance Level for Version-Current Packages (VND $<$ 1.0, n = 9,918)}
\label{tab:rq2-malta-low-vl}
\centering
\begin{tabular}{llccc}
\toprule
\textbf{Level} & \textbf{$S^{(100)}_{final}$} & \textbf{n} & \textbf{\%} & \textbf{Median TL (days)} \\
\midrule
Sustained & 80--100 & 1,000 & 10.1\% & 10 \\
Stable & 60--79 & 1,143 & 11.5\% & 48 \\
Declining & 40--59 & 1,608 & 16.2\% & 104 \\
Probable Abandonment & 20--39 & 940 & 9.5\% & 174 \\
Effective Abandonment & 0--19 & 5,227 & 52.7\% & 3652 \\
\bottomrule
\end{tabular}
\end{table}

Table~\ref{tab:rq2-malta-low-vl} and Figure~\ref{fig:rq2-malta-low-vl} examine the full MALTA score distribution among the \rqtwoLowVLN{} version-current packages, using the interpretive scale defined in
Section~\ref{desc:final-score-interpretation}.

Figure~\ref{fig:rq2-malta-low-vl}(a) overlays the $S_{final}$ distribution of version-current packages against the rest of the dataset. While the distributions share a similar shape, the version-current population retains a substantial left tail below the MALTA risk thresholds, confirming that version currency alone does not indicate healthy maintenance.

The stacked-bar comparison in Figure~\ref{fig:rq2-malta-low-vl}(b) quantifies this directly: among version-current packages, \rqtwoLowVLHighMaltaPct{}\% fall into MALTA High Risk comparable to the \rqtwoAllHighMaltaPct{}\% rate across the full dataset. Version currency thus provides almost no protective signal against maintenance decline as measured by MALTA.

These findings directly motivate the risk reclassification analysis in Section~\ref{sec:results-rq3}; a substantial fraction of the version-current population would be identified as at-risk if maintenance signals were incorporated into a TL assessment.

\begin{tcolorbox}[
    enhanced,attach boxed title to top center={yshift=-3mm,yshifttext=-1mm},
    title=Answer to RQ2,fonttitle=\bfseries,
    boxed title style={size=small}]
Sedentary packages exhibit \rqtwoTimeLagMedianRatio{} a higher median Time Lag than Very Active packages. Among version-current packages, \rqtwoLowVLAbandonedCombPct{}\% show indicators of abandonment under MALTA, revealing a hidden population decline invisible to Version Lag alone.
\end{tcolorbox}

\subsection{\textbf{RQ3:} Incorporating Maintenance Decline}\label{sec:results-rq3}

We compared Version Lag risk (based on Version Number Delta between Debian releases) with MALTA-informed risk (based on $S_{final}$ thresholds from Section~\ref{desc:final-score-interpretation}) across all \rqthreeNPackages{} packages, applying the same Time Lag capping as in RQ2.

\subsubsection{Risk Reclassification Analysis}\label{sec:rq3-reclassification}

\begin{figure*}[htp]
    \centering
    \includegraphics[width=\linewidth]{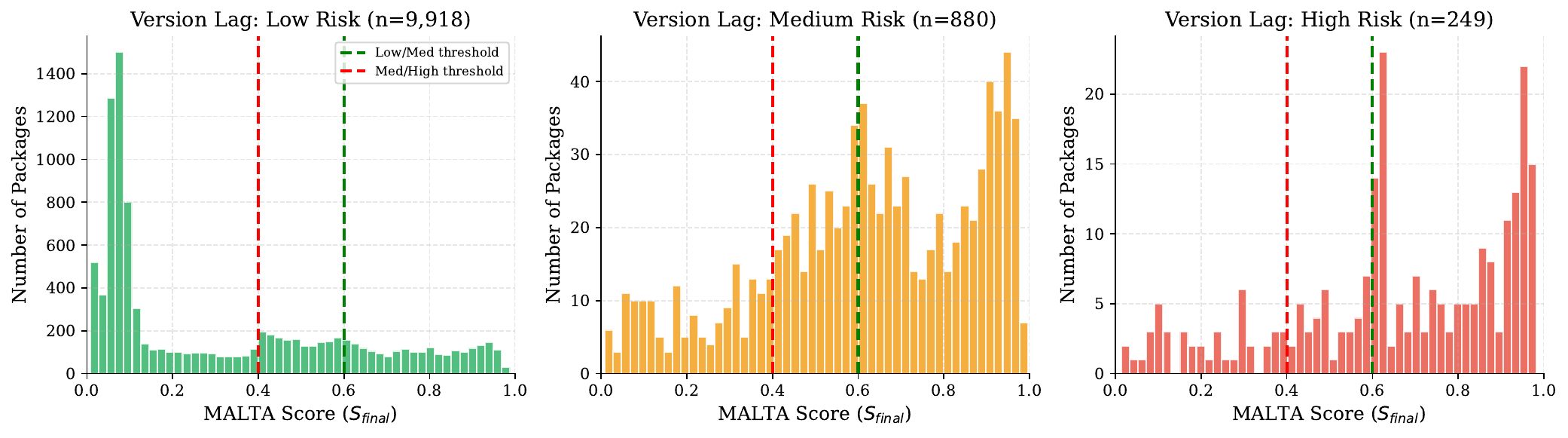}
    \caption{MALTA Score Distribution by Version Lag Risk Category. Vertical lines indicate MALTA thresholds.}
    \label{fig:rq3-risk-reclass}
\end{figure*}

\begin{table}[t]
\scriptsize
\centering
\caption{Cross-tabulation of Version Lag vs. MALTA Risk Classifications (n = 11,047)}
\label{tab:rq3-crosstab}
\begin{tabular}{l|rrr|r}
\toprule
\multirow{2}{*}{\textbf{Version Lag Risk}} & \multicolumn{3}{c|}{\textbf{MALTA Risk}} & \multirow{2}{*}{\textbf{Total}} \\
& Low & Medium & High & \\
\midrule
Low Risk & 2,143 (21.6\%) & 1,608 (16.2\%) & 6,167 (62.2\%) & 9,918 \\
Medium Risk & 500 (56.8\%) & 217 (24.7\%) & 163 (18.5\%) & 880 \\
High Risk & 166 (66.7\%) & 39 (15.7\%) & 44 (17.7\%) & 249 \\
\midrule
Total & 2,809 & 1,864 & 6,374 & 11,047 \\
\bottomrule
\end{tabular}
\end{table}

The two risk classifications are statistically associated but show extensive disagreement (Table~\ref{tab:rq3-crosstab}, Figure~\ref{fig:rq3-risk-reclass}). A chi-square test confirms that the two classifications are not independent ($\chi^2 = \rqthreeChiSq{}$, $df = 4$, $p < 0.001$, Cram\'{e}r's $V = \rqthreeCramersV{}$).

The cross-tabulation reveals an \emph{inverse} relationship between version currency and maintenance health. Among VL-Low packages (\rqthreeLagLowTotal{}, VND $<$ \rqtwoLowVNDThreshold{}), only \rqthreeLagLowMaltaLowPct{}\% are confirmed as Low Risk by MALTA, while \textbf{\rqthreeLagLowMaltaHighPct{}\%} (\rqthreeLagLowMaltaHighN{}) are classified as High Risk. Conversely, among VL-High packages, \rqthreeLagHighMaltaLowPct{}\% are MALTA Low Risk and only \rqthreeLagHighMaltaHighPct{}\% are MALTA High Risk, a \rqthreeMaltaHighRateRatio{}$\times$ lower rate than the VL-Low population. The relevant comparison is across VL categories, where packages with large version deltas are overwhelmingly maintained while version-current packages are not.

This bidirectional disagreement has a clear explanation. Packages with high Version Lag have active upstreams that continue producing releases, resulting in large version deltas that \emph{signal} upstream health rather than downstream neglect. Packages with low Version Lag include a substantial population of abandoned projects whose version currency is a frozen artifact of upstream cessation, not evidence of maintenance. The histograms in Figure~\ref{fig:rq3-risk-reclass} illustrate this pattern: the MALTA score distribution among VL-Low packages is heavily left-skewed, with the majority falling below the Medium Risk threshold.

\subsubsection{Discordant Classification Analysis}\label{sec:rq3-discordant}

\begin{table}[t]
\centering
\caption{Characteristics of Discordant vs. Concordant Classifications}
\label{tab:rq3-discordant}
\begin{tabular}{lrr}
\toprule
\textbf{Metric} & \textbf{Discordant} & \textbf{Concordant} \\
\midrule
Count & 6,167 & 2,143 \\
Percentage of Dataset & 55.8\% & 19.4\% \\
Mean MALTA Score ($S_{final}$) & 0.109 & 0.786 \\
Mean DAS Score ($S_{dev}$) & 0.055 & 0.787 \\
Mean Time Lag (days) & 2019 & 36 \\
Archived Repositories & 9.8\% & 0.0\% \\
Sedentary (PVAC) & 81.8\% & 40.5\% \\
\bottomrule
\end{tabular}
\end{table}

Table~\ref{tab:rq3-discordant} characterizes the \rqthreeFalseHealthyN{} discordant packages, those classified as Low Risk by Version Lag but High Risk by MALTA, and compares them with the \rqthreeTrulyHealthyN{} concordant packages (Low Risk under both methods). The two populations differ sharply across every maintenance indicator (Mann-Whitney $U$, $p < 0.001$ for all comparisons).

Discordant packages exhibit a mean MALTA score of \rqthreeFalseHealthyMALTAMean{} compared to \rqthreeTrulyHealthyMALTAMean{} for concordant packages, reflecting near-zero development activity. Their mean Time Lag of \rqthreeFalseHealthyTimeLagMean{} days is \rqthreeTimeLagRatio{}$\times$ longer than the \rqthreeTrulyHealthyTimeLagMean{}-day mean for concordant packages, and \rqthreeFalseHealthyArchivedPct{}\% of their repositories have been explicitly archived (vs.\ \rqthreeTrulyHealthyArchivedPct{}\%). Furthermore, \rqthreeFalseHealthySedentaryPct{}\% are classified as Sedentary by PVAC, confirming that these packages have experienced long-term maintenance cessation that Version Lag fails to detect.

\begin{tcolorbox}[
    enhanced,attach boxed title to top center={yshift=-3mm,yshifttext=-1mm},
    title=Answer to RQ3,fonttitle=\bfseries,
    boxed title style={size=small}]
Version Lag and MALTA risk classifications are inversely related. Version-current packages (VL-Low) are \rqthreeMaltaHighRateRatio{}$\times$ more likely to be MALTA-High than packages with large version deltas (\rqthreeLagLowMaltaHighPct{}\% vs.\ \rqthreeLagHighMaltaHighPct{}\%), because version currency in abandoned packages is a frozen artifact of upstream cessation. Only \rqthreeLagLowMaltaLowPct{}\% of VL-Low packages are confirmed as Low Risk by both methods.
\end{tcolorbox}

\subsection{Sensitivity Analysis}\label{sec:sensitivity}

\begin{table*}[t]
\caption{Sensitivity Analysis: Classification Performance and Risk Distribution Under Alternative Parameter Configurations}
\label{tab:sensitivity}
\centering
\footnotesize
\begin{tabular}{llccccc}
\toprule
\textbf{Dimension} & \textbf{Configuration} & \textbf{AUC$_\text{act}$} & \textbf{AUC$_\text{arch}$} & \textbf{\% Low} & \textbf{\% High} & \textbf{Agree} \\
\midrule
Component Weights & \textbf{Default (55/35/10)} & \textbf{0.783} & \textbf{0.944} & \textbf{25.4} & \textbf{57.7} & \textbf{100.0} \\
 & Equal (33/33/34) & 0.760 & 0.984 & 25.4 & 54.2 & 90.1 \\
 & DAS-heavy (70/20/10) & 0.796 & 0.930 & 26.6 & 61.8 & 87.2 \\
 & MRS-heavy (20/70/10) & 0.720 & 0.984 & 30.7 & 63.5 & 80.1 \\
 & Balanced (40/40/20) & 0.765 & 0.972 & 25.5 & 57.2 & 92.5 \\
 & DAS-only (100/0/0) & 0.803 & 0.743 & 27.1 & 64.5 & 78.9 \\
\midrule
DAS Decay ($\tau$) & \textbf{Default ($\tau$=180)} & \textbf{0.783} & \textbf{0.944} & \textbf{25.4} & \textbf{57.7} & \textbf{100.0} \\
 & Short decay ($\tau$=90) & 0.778 & 0.966 & 20.1 & 62.9 & 89.4 \\
 & Long decay ($\tau$=365) & 0.778 & 0.929 & 31.1 & 53.4 & 90.1 \\
\midrule
MRS Timeliness ($T_{ref}$) & \textbf{Default ($T_{ref}$=180)} & \textbf{0.783} & \textbf{0.944} & \textbf{25.4} & \textbf{57.7} & \textbf{100.0} \\
 & Short window ($T_{ref}$=90) & 0.781 & 0.943 & 23.9 & 58.9 & 97.2 \\
 & Long window ($T_{ref}$=365) & 0.789 & 0.946 & 28.0 & 55.8 & 95.5 \\
\midrule
RMVS Parameters & \textbf{Default ($\alpha$=0.7, equal $\beta$)} & \textbf{0.783} & \textbf{0.944} & \textbf{25.4} & \textbf{57.7} & \textbf{100.0} \\
 & Mild archival ($\alpha$=0.5) & 0.783 & 0.935 & 25.4 & 57.7 & 100.0 \\
 & Severe archival ($\alpha$=0.9) & 0.783 & 0.949 & 25.4 & 57.7 & 100.0 \\
 & Popularity-weighted $\beta$ & 0.785 & 0.937 & 25.8 & 57.7 & 98.9 \\
\midrule
MALTA Risk Thresholds & \textbf{Default (0.60/0.40)} & \textbf{0.783} & \textbf{0.944} & \textbf{25.4} & \textbf{57.7} & \textbf{100.0} \\
 & Strict (0.70/0.50) & 0.783 & 0.944 & 18.2 & 66.6 & 83.8 \\
 & Lenient (0.50/0.30) & 0.783 & 0.944 & 33.4 & 53.1 & 87.4 \\
 & Narrow (0.55/0.45) & 0.783 & 0.944 & 29.7 & 62.5 & 91.0 \\
 & Wide (0.70/0.30) & 0.783 & 0.944 & 18.2 & 53.1 & 88.1 \\
\midrule
Version Lag Thresholds & \textbf{Default (1.0/3.0)} & \textbf{0.778} & \textbf{0.600} & \textbf{89.8} & \textbf{2.3} & \textbf{100.0} \\
 & Strict (0.5/2.0) & 0.778 & 0.600 & 80.3 & 4.2 & 88.5 \\
 & Lenient (2.0/5.0) & 0.778 & 0.600 & 95.8 & 1.1 & 92.9 \\
 & Narrow (1.0/2.0) & 0.778 & 0.600 & 89.8 & 4.2 & 98.0 \\
 & Wide (0.5/5.0) & 0.778 & 0.600 & 80.3 & 1.1 & 89.3 \\
\bottomrule
\end{tabular}
\end{table*}

Table~\ref{tab:sensitivity} reports classification performance and risk-label stability across six parameter dimensions. Component weight variations confirm that DAS is the strongest individual predictor (AUC$_\text{act}$=0.803 when used alone), but omitting the other two signals reduces archived-repository detection (AUC$_\text{arch}$=0.743), justifying the multi-component design. The DAS decay constant $\tau$ and MRS timeliness reference $T_{ref}$ each show modest sensitivity when varied independently (agreement $\geq$89\%), with no alternative improving both AUC metrics simultaneously. RMVS parameters are effectively inert: varying $\alpha$ from 0.5 to 0.9 leaves risk classifications unchanged ($\geq$99.9\% agreement), and non-equal $\beta$ weights produce 98.9\% agreement, confirming that the 10\% RMVS weight appropriately limits the influence of popularity metadata.
\section{Discussion}\label{sec:discussion}

This section interprets our empirical findings, discusses their implications for both practitioners and researchers, and situates MALTA within the broader landscape of TL and software abandonment research.  Importantly, MALTA's discriminative ability is not an artifact of shared signal overlap, but a complementary perspective to a project's health. The PVAC-based ground truth used in Section~\ref{sec:rq1-classification-performance} relies on versioning activity, while DAS measures commit activity, conceptually related but operationally distinct signals. To guard against circular validation, we conducted an independent evaluation against explicitly archived repositories (Section~\ref{sec:rq1-archive-validation}), removing the archival flag from MALTA's score computation. The modified score $S_{final}^{-A}$ achieved an AUC of \rqoneAUCNoArchFinal{} for archived prediction, and \rqoneArchHighRiskPct{}\% of archived repositories were classified as High Risk. This convergence across two independent ground truths, one based on versioning patterns, the other on deliberate maintainer decisions, strengthens confidence that MALTA captures a genuine maintenance construct rather than a measurement artifact.

The magnitude of risk reclassification when MALTA signals are incorporated (Table~\ref{tab:rq3-crosstab}) underscores the practical significance of maintenance-aware TL assessment. Of \rqthreeLagLowTotal{} packages classified as Low Risk by Version Lag alone, \rqthreeLagLowMaltaHighPct{}\% (\rqthreeLagLowMaltaHighN{} packages) were reclassified as High Risk. For organizations relying on Version Lag as their primary dependency health indicator, the implication is stark: their risk assessment systematically overlooks the most vulnerable packages while flagging actively maintained ones whose rapid upstream development produces large version deltas. Using multiple lag functions has been confirmed in prior research~\cite{zeroualiMultidimensionalAnalysisTechnical2021}, strengthening our results.

This finding is not an artifact of MALTA's overall classification distribution: packages with \emph{high} Version Lag are \rqthreeLagHighMaltaLowPct{}\% MALTA-Low, demonstrating that large version deltas are predominantly a signal of active upstream development rather than dependency risk. The concentration of MALTA-High classifications within the version-current population confirms that Version Lag's blind spot is not uniformly distributed but specifically targets abandoned packages whose version currency masks upstream cessation.

It is important to note that MALTA is not a binary classifier but a continuous score that can be configured at different levels to balance precision and recall according to specific risk tolerance. MALTA has been designed to be complementary to existing lag functions, not a replacement. For practitioners, the key takeaway is that relying solely on Version Lag can lead to a false sense of security, while incorporating MALTA provides a more nuanced and accurate picture of dependency health. Its value emerges most clearly in the context of version-current packages, where it can distinguish between those that are actively maintained and those that are effectively abandoned.

We have identified two new categories of \textbf{lag functions} that capture the maintenance state of a package with respect to its ability to be resolved: \textit{resolvable lag} (actively maintained packages with update opportunities) and \textit{terminal lag} (abandoned packages with no update path). This extends the existing taxonomy of lag functions~\cite{panterTechnicalLagLatent2026a} and enables more robust multidimensional assessment of TL.

\subsection{Implications for Practitioners}

\subsubsection{Dependency Selection and Monitoring}

Version Lag remains valuable for identifying \textit{update opportunities} in actively maintained packages, but should be supplemented with maintenance activity indicators to detect \textit{abandonment risk}.

\subsubsection{Distribution Maintainers}

For maintainers of LAPM (e.g., Debian, Fedora), MALTA provides actionable intelligence for package lifecycle decisions. Distributions such as Debian have a list of ``orphaned'' packages\footnote{https://www.debian.org/devel/wnpp/orphaned} without a current maintainer. MALTA can further augment this list by identifying at-risk packages before they reach that point, enabling proactive interventions such as seeking new maintainers, merging with active forks, or planning for migration.

\subsection{Implications for Researchers}

Reconceptualizing TL as either resolvable or terminal enables more precise modeling and risk assessment. The traditional formulation of TL~\cite{gonzalez-barahonaTechnicalLagSoftware2017} implicitly assumes that lag can be resolved through updates, an assumption that breaks down for abandoned packages.

\subsection{Limitations of the Current Approach}

Beyond the threats to validity discussed in Section~\ref{sec:threats}, several limitations merit discussion:

\subsubsection{Platform Dependency}
Our current implementation relies on GitHub-specific signals (stars, forks, pull requests). While these signals have analogs on other platforms, the specific thresholds and distributions may differ. The saturation constants ($K_x$) computed from our Debian/GitHub dataset may not transfer directly to other ecosystems.

\subsubsection{Temporal Validity}
MALTA scores reflect the maintenance state at the time of computation and may become stale. A package classified as ``Sustained Maintenance'' could experience a rapid decline following a maintainer's departure. We recommend periodic re-computation (e.g., quarterly) for dependencies in active use.

\subsubsection{Aggregate vs. Granular Risk}
MALTA produces a single aggregate score that may mask heterogeneous risk profiles. A package with excellent DAS but poor MRS presents different risks than one with the inverse pattern. Future work could provide component-level risk profiles for more nuanced assessment.

\section{Threats to Validity}\label{sec:threats}

We organize threats to validity following the framework of Wohlin et al.~\cite{wohlinExperimentationSoftwareEngineering2012}, addressing construct, internal, and external.

\subsection{Construct Validity}

We used the Package Version Activity Categorizer (PVAC) to stratify packages into maintenance categories for analysis. Packages with non-standard versioning schemes or those that backport fixes without version bumps may be misclassified. However, PVAC has been validated in prior work~\cite{panterPVACPackageVersion2025} and provides a reproducible, objective baseline independent of MALTA's own signals. The DAS excludes commits deemed non-substantive; our filtering heuristics may incorrectly exclude meaningful maintenance activity or include trivial changes. The MRS may not hold for projects that primarily receive contributions through other channels. We assume that projects that list GitHub as their upstream version control and project homepage are primarily maintained using GitHub-provided tools (Pull Requests, issues, etc.), which may not hold for all projects. Some projects may simply use GitHub for mirroring, for visibility only; for example, the Linux kernel\footnote{https://github.com/torvalds/linux}. Additionally, our archival validation relies on the assumption that archived repositories are representative of abandoned projects rather than the result of software evolution or other factors.

\subsection{Internal Validity}

Our dataset includes only Debian packages with identifiable GitHub upstreams, representing approximately 40\% of total Debian packages. Packages hosted on other platforms (GitLab, self-hosted repositories, SourceForge) or without public version control may differ systematically in maintenance patterns. Additionally, packages selected for Debian inclusion have already passed distribution maintainer review, potentially biasing our sample toward higher-quality projects than the broader open-source population.

\subsection{External Validity}

 Our empirical evaluation focused exclusively on the Debian ecosystem, which has distinct characteristics: long release cycles (approximately 2 years), conservative update policies prioritizing stability, and a mature package maintainer community. Findings may not generalize to rolling-release distributions (e.g., Arch Linux, Fedora Rawhide), language-specific package managers (e.g., npm, PyPI, Maven), or commercial software ecosystems. The relationship between TL and maintenance state may differ substantially in ecosystems with more aggressive update policies or different governance structures. However, if MALTA works in Debian, it is likely to work in faster-moving ecosystems.

\section{Conclusion and Future Work}\label{sec:conclusion}

TL metrics have become essential tools for assessing dependency health in modern software ecosystems. However, existing approaches share a critical blind spot: they cannot distinguish between packages that are current because they are well-maintained and those that appear current because upstream development has ceased. This limitation leads to systematic underestimation of risk for abandoned dependencies, precisely the packages that pose the greatest long-term sustainability and security concerns.

\subsection{Future Work}

Each ecosystem has unique characteristics that may influence the relationship between maintenance signals and TL. While every attempt has been made to design MALTA to be platform-agnostic, validating its applicability to other LAPMs, such as \textit{dnf}, would strengthen confidence in its generalizability. LSPMs such as \textit{npm} and \textit{PyPI} exhibit different contribution patterns and release cadences, which may affect the relevance of certain signals. Additionally, LAPMs that use a rolling release model (e.g., Arch Linux) may require different approaches to version-based signals because there is no fixed release cadence to anchor version-based metrics for users.

Currently, MALTA does not incorporate direct measures of security risk, such as known vulnerabilities. Future work could explore integrating vulnerability data from sources like the National Vulnerability Database (NVD)\footnote{https://nvd.nist.gov} or GitHub's Security Advisories\footnote{https://github.com/advisories} to provide a more comprehensive risk assessment that combines maintenance status with security posture. Additionally, MALTA focuses on estimating the likelihood of abandonment but does not account for the possibility that projects may have moved maintenance to a private repository and simply mirrored the public repository to GitHub.

Finally, detecting packages that use GitHub solely for mirroring, rather than as their primary development platform, is an open challenge. Future work could leverage LLMs to read project documentation and commit messages to identify signs of mirroring-only usage, such as references to an alternative primary repository or a lack of typical GitHub maintenance activity.

\subsection{Closing Remarks}

This paper introduces MALTA, a scoring framework that improves TL methods and metrics to distinguish between \textit{resolvable lag} and \textit{terminal lag} caused by upstream abandonment. By incorporating maintenance signals such as development activity, maintainer responsiveness, and repository metadata, MALTA can help practitioners identify high-risk dependencies that would otherwise be misclassified as low risk. Our empirical evaluation of GitHub-hosted packages demonstrates the need to add maintenance-aware TL metrics to the TL toolbox, enabling more accurate and actionable assessments of dependency health in open-source ecosystems.

\bibliographystyle{IEEEtran}
\bibliography{paper}
\balance
\end{document}